\documentclass[aps,prl,twocolumn,superscriptaddress,nobalancelastpage,longbibliography]{revtex4-2}
\usepackage{amsmath}
\usepackage{amsfonts}
\usepackage{amssymb}
\usepackage{graphicx}
\usepackage{xcolor}
\usepackage[colorlinks=True,citecolor=myOrange,linkcolor=myRed,urlcolor=myOrange]{hyperref}
\usepackage{hypcap}
\usepackage{braket}
\usepackage{stmaryrd}

\newcommand\eq[1]{\begin{align}#1\end{align}}
\newcommand\npeff{N_q^{\mathrm{eff}}}
\newcommand{\pmax}[1]{P_{\mathrm{max}}\left(#1\right)}
\newcommand{\cmax}[1]{C_{\mathrm{max}}\left(#1\right)}

\definecolor{myBlue}{RGB}{31,119,180}
\definecolor{myOrange}{RGB}{255,127,14}
\definecolor{myGreen}{RGB}{44,160,44}
\definecolor{myRed}{RGB}{214,39,40}
\definecolor{myPurple}{RGB}{148,103,189}

\makeatletter
\def\p@figure{\color{myBlue}}
\def\p@equation{\color{myRed}}

\makeatother

\begin{document}

\title{
How periodic driving stabilises and destabilises Anderson localisation on random trees
}

\author{Sthitadhi Roy}
\affiliation{Rudolf Peierls Centre for Theoretical Physics, Clarendon Laboratory,
Oxford University, Parks Road, Oxford OX1 3PU, United Kingdom}
\affiliation{Physical and Theoretical Chemistry, Oxford University,
South Parks Road, Oxford OX1 3QZ, United Kingdom}
\author{Roderich Moessner}
\affiliation{Max-Planck-Institut f\"ur Physik komplexer Systeme, N\"othnitzer Stra{\ss}e 38, 01187 Dresden, Germany}
\author{Achilleas Lazarides}
\affiliation{Interdisciplinary Centre for Mathematical Modelling and Department of Mathematical Sciences, Loughborough University, Loughborough, Leicestershire LE11 3TU, United Kingdom}

\begin{abstract}
Motivated by the link between Anderson localisation on high-dimensional graphs and many-body localisation, we study the effect of periodic driving on Anderson localisation on random trees. The time dependence is eliminated in favour of an extra dimension, resulting in an extended graph wherein the disorder is correlated along the new dimension. The extra dimension increases the number of paths between any two sites and allows for interference between their amplitudes. We study the localisation problem within the forward scattering approximation (FSA) which we adapt to this extended graph. At low frequency, this favours delocalisation as the availability of a large number of extra paths dominates. By contrast, at high frequency, it stabilises localisation compared to the static system. These lead to a regime of re-entrant localisation in the phase diagram. Analysing the statistics of path amplitudes within the FSA, we provide a detailed theoretical picture of the physical mechanisms governing the phase diagram. 
\end{abstract}

\maketitle

Localisation in quantum systems~\cite{anderson1958absence,basko2006metal,gornyi2005interacting,oganesyan2007localisation,znidaric2008many,nandkishore2015many} renders invalid the framework of statistical mechanics and paves the way to novel quantum phases and order in excited eigenstates~\cite{huse2013localisation}. Perhaps the most remarkable examples of these are the discrete time crystals in a periodically-driven (Floquet) setting~\cite{khemani2016phase,else2016floquet,yao2017discrete,moessner2017equilibration}. This spatiotemporal ordering is an inherently non-equilibrium phenomenon~\cite{watanabe2015absence} apparently impossible in systems governed by static Hamiltonians~\cite{khemani2019brief}. While ergodic Floquet systems generically heat up to a featureless ``infinite-temperature'' state~\cite{lazarides2014equilibrium,dalessio2014long}, breaking ergodicity robustly via disorder-induced localisation prevents this heat death~\cite{ponte2015many,lazarides2015fate,ponte2015periodically,abanin2016theory}. Understanding the mechanisms that might stabilise or destabilise localisation in the presence of periodic driving is therefore a question of immanent interest.

\begin{figure}
\includegraphics[width=\linewidth]{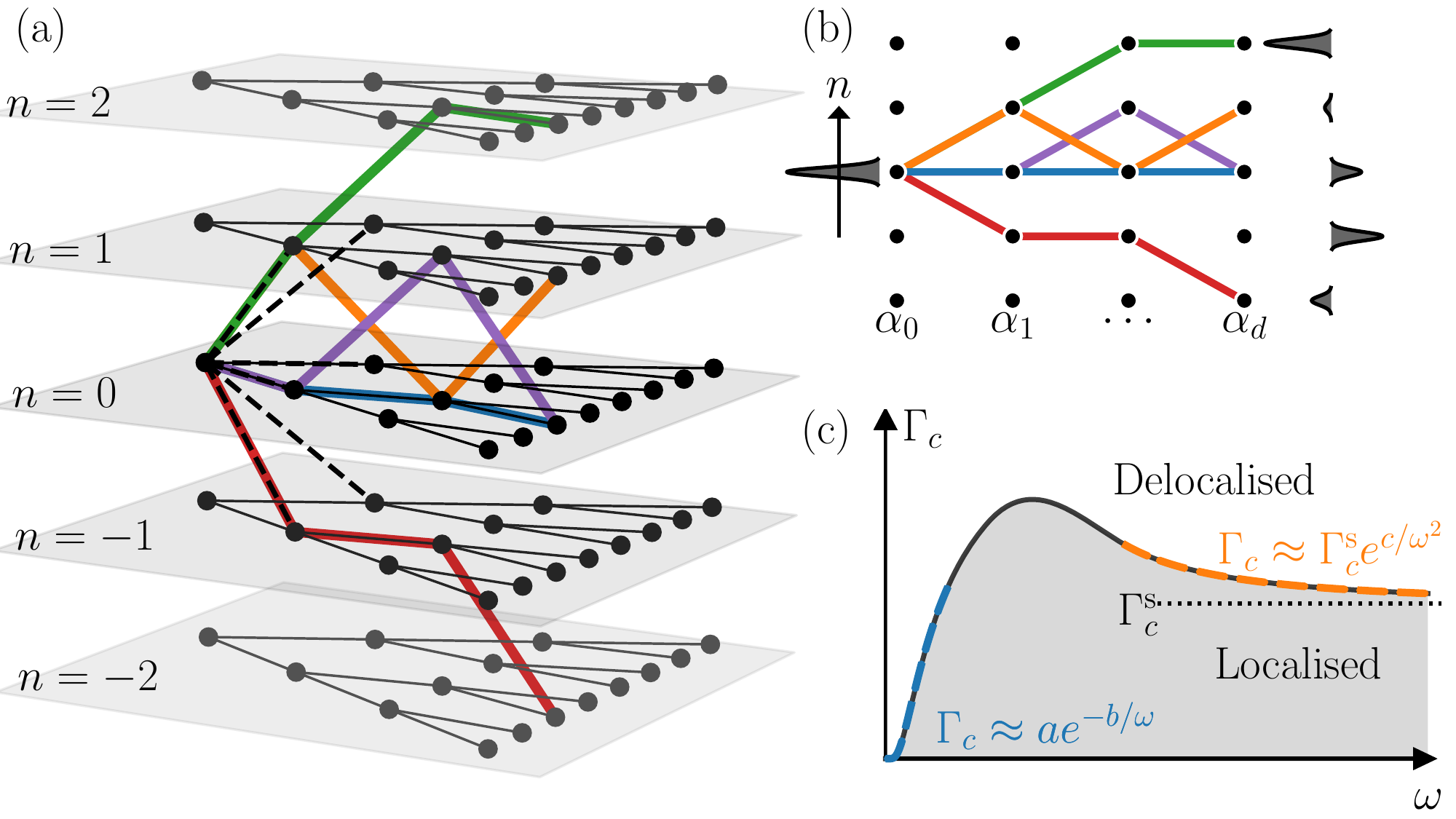}
\caption{(a) The Shirley picture for a driven tight-binding model on a tree. The different copies of the tree are indexed by $n$. The blue line shows the \emph{static} path between the root site and another, say $\alpha_0$ and $\alpha_d$, whereas the different coloured lines show some of the driving-induced paths between the root site and the same site's copy, $\ket{\alpha_d,n}$, on other Shirley layers. (b) Schematic of the FSA amplitude on sites $\ket{\alpha_d,n}$ starting from $\ket{\alpha_0,0}$. The amplitudes on the former can be both positive and negative so they can interfere to give the total amplitude $\psi_{\alpha_d}\propto\sum_{n}\psi_{\alpha_d,n}$.  (c) Summary of the localisation phase diagram in the $(\Gamma,\omega)$ plane for the model in (a). At high frequencies, driving favours localisation and enhances the critical hopping compared to the static case, $\Gamma_c^\mathrm{s}$, whereas at low frequencies, it favours delocalisation and $\Gamma_c$ decreases with decreasing $\omega$.}
\label{fig:shirley}
\end{figure}

While the problem of many-body localisation (driven or otherwise) can also be equivalently viewed as one of single-particle localisation on the high-dimensional Fock-space graph, strong correlations in the latter render it qualitatively different from conventional Anderson localisation on such graphs~\cite{roy2020fock,roy2020localisation}.
Nevertheless, a natural step towards a theoretical understanding of the mechanism governing localisation in Floquet systems is to analytically study a simpler problem in a more controlled setting -- the fate of Anderson transitions and localisation on high-dimensional graphs.
We, therefore, study Floquet tight-binding models defined on random high-dimensional graphs which are generally tree-like locally.
In particular, the model we consider is described by a time-periodic (with period $T\equiv2\pi/\omega$) Hamiltonian,
\eq{
H(t) = H_0 + H_1\cos(\omega t)\,,
\label{eq:Ht}
}
with $H_0=H_\mathrm{hop} + H_\mathrm{dis}$ and $H_1=H_\mathrm{hop}$. $H_\mathrm{hop}$ describes the hopping  while $H_\mathrm{dis}$ the onsite disorder, 
\eq{
    H_\mathrm{hop} = \Gamma\sum_{\braket{\alpha,\beta}}\ket{\alpha}\bra{\beta}+\mathrm{H.c.};\; H_\mathrm{dis}=\sum_\alpha\epsilon_\alpha\ket{\alpha}\bra{\alpha}\,,
    \label{eq:H-hop-dis}
}
respectively where $\braket{\alpha,\beta}$ denotes links between pairs of sites on the graph, and the random potentials are drawn from a distribution with width $W$.
Undriven cousins of the model in Eq.~\ref{eq:Ht}, \textit{i.e.} disordered tight-binding models on trees and random regular graphs have long served as archetypes for localisation transitions and related phenomena on high-dimensional graphs, allowing for a remarkable amount of theoretical progress~\cite{abou-chacra1973self,chalker1990anderson,evers2008anderson,luca2014anderson,altshuler2016multifractal,tikhonov2016anderson,garciamata2017scaling,sonner2017multifractality,biroli2018delocalization,kravtsov2018nonergodic,tikhonov2019critical,savitz2019anderson,garciamata2020two,tarzia2020manybody,biroli2017delocalized,biroli2020anomalous}.

Following the approach pioneered by Shirley~\cite{shirley1965solution,sambe1973steady}, the periodic time-dependence of a Hamiltonian of the form \eqref{eq:Ht} can be eliminated in favour of an extra dimension. Generalising the Forward Scattering Approximation (FSA)~\cite{pietracaprina2016forward} to this Shirley picture, we focus on the additional paths (compared to the static case) the extra dimension allows for between any two physical sites on the graph. The interplay between these extra paths becoming available and their interference results in an amplification or attenuation of the amplitude relative to that of the static one. We provide analytical estimates for the locations of Anderson transitions in the hopping strength--frequency plane, Fig.~\ref{fig:shirley}. 
The phase diagram exhibits a characteristic re-entrant localised portion.

The remainder of this paper is organised as follows. We first recapitulate the Shirley formalism, showing how the time dependence is eliminated in favour of an extra dimension, then generalise the FSA to this situation. We then provide numerical evidence for the localisation phase diagram (Fig.~\ref{fig:shirley}(c)), followed by analytical arguments elucidating the mechanisms underpinning it. 

The problem in the Shirley picture maps to a new undriven system on an infinite ladder where each rung corresponds to a copy of the static part of the Hamiltonian $H_0$ with the $n^\mathrm{th}$ rung shifted in energy by $n\omega$, and the driving Hamiltonian, $H_1$, mediating hopping between the rungs. Thus the new system lives on an extended graph with one additional dimension.
This time-independent Hamiltonian, for the model described by Eqs.~\ref{eq:Ht} and \ref{eq:H-hop-dis}, takes the form
\begin{equation}
\begin{split}
    H_F = &\Gamma\sum_{\braket{\alpha,\beta},n}\sum_{s=0,\pm 1}[\ket{\alpha,n}\bra{\beta,n+s}+\mathrm{H.c}]\\&+\sum_{\alpha,n}(\epsilon_\alpha+n\omega)\ket{\alpha,n}\bra{\alpha,n}\,,
    \label{eq:H-shirley}
\end{split}
\end{equation}
where $\ket{\alpha,n}=\ket{\alpha}\otimes\ket{n}$ denotes a state localised on physical site $\alpha$ and Shirley rung $n$.
Solutions of the time-dependent Schr\"odinger equation $\left(i\partial_t-H(t)\right)\ket{\phi(t)}=0$ can be written 
as $\ket{\phi_a(t)}=e^{i\Omega_a t}\sum_{n}\ket{\varphi_a^{(n)}}e^{in\omega t}$, 
where 
$\ket{\varphi_a}=\sum_{n=-\infty}^\infty\ket{\varphi_a^{\scriptscriptstyle{(n)}}}\otimes\ket{n}$ 
is an eigenstate of the time-independent 
Hamiltonian \eqref{eq:H-shirley} with eigenvalue $\Omega_a$ and $\ket{\varphi_a^{\scriptscriptstyle{(n)}}}$ is a state living in a single rung. 

Physically, the hopping between different rungs corresponds to the system gaining or losing energy in integer multiples of $\omega$.
Thus additional processes, or in the time-independent picture, additional paths are made available to hop and delocalise away from a site $\alpha$ by the driving, see Fig.~\ref{fig:shirley} for a visual representation. 
The question of localisation for the driven system \eqref{eq:Ht} is now mapped onto one for the time-independent system \eqref{eq:H-shirley} with one extra dimension, which we will study within the FSA. 

\begin{figure*}
\includegraphics[width=\linewidth]{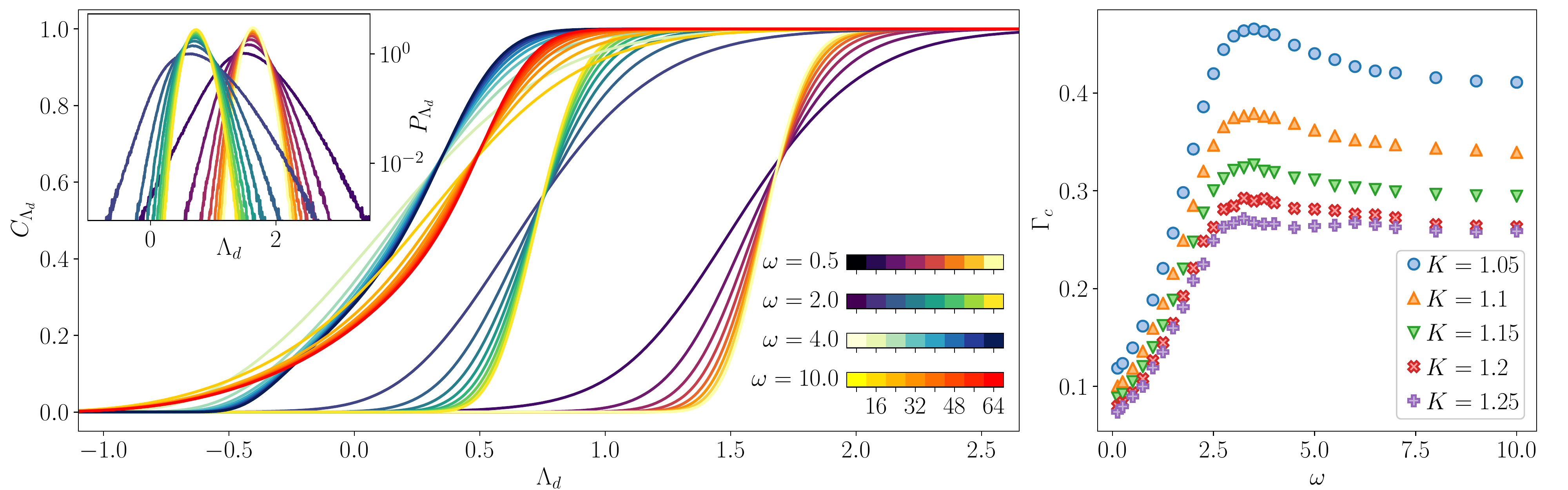}
\caption{(a) The cumulative distribution of the rescaled FSA amplitudes, $C(\Lambda_d)$ (see Eqs.~\ref{eq:deloc-crit} and \ref{eq:FSA-critical}) for values of $d=8,16,\cdots,64$ and four different values of $\omega$ labelled by the four colour-maps. The crossing points of the curves indicate the $\ln(W/\Gamma)_c$ for the corresponding $\omega$. The inset shows the distributions $P(\Lambda_d)$ for $\omega=0.5$ and $\omega=2$ which get sharper with increasing $d$. Results are for $K=1.05 $ and the disordered potential was drawn from a standard Normal distribution $\epsilon_\alpha\sim \mathcal{N}(0,W=1)$. (b) The critical $\Gamma_c/W$ as a function of $\omega/W$ for different values of $K$ obtained from the crossing of $C(\Lambda_d)$ exemplified in (a) and obtained using Eq.~\ref{eq:PLambda-K}. Note the non-monotonicity in $\Gamma_c/W$ as a function of $\omega/W$. Statistics for all data were obtained over $5\times 10^6$ realisations. }
\label{fig:fsa}
\end{figure*}

We begin by briefly describing the FSA for a static system. 
For a state initially (in the absence of hopping) localised at $\alpha_0$, the FSA estimates the amplitude on some other site, $\alpha_d$ at distance $d$, upon perturbatively including the effects of hopping by summing over all the shortest paths between the two sites. It is thus a stability analysis of the localised phase at $\Gamma=0$ for $\Gamma\neq 0$~\cite{pietracaprina2016forward}. The amplitude $\psi_{\alpha_d}$ is
\eq{
\psi_{\alpha_d} = \sum_{p:\alpha_0\shortrightarrow\alpha_d}\prod_{\beta\in p}\frac{\Gamma}{\epsilon_{\beta}-\epsilon_{\alpha_0}}\,,
\label{eq:fsa-stat}
}
where the sum is over all the shortest paths from $\alpha_0$ to $\alpha_d$.
Breakdown of localisation is indicated by the probability (over disorder realisations) of the state spreading to at least one site at arbitrarily large distance $d$ approaching unity.
Defining $\vert\psi_d\vert^2=\max_{\alpha_d}\vert\psi_{\alpha_d}\vert^2$, the maximum amplitude over all sites at a given distance, the delocalisation criterion is
\eq{
\lim_{d\to\infty}P\left(\frac{\ln\vert\psi_d\vert^2}{2d}>-\xi^{-1}\right)\to 1\,,
\label{eq:deloc-crit}
}
where $\xi$ is the localisation length. 
Defining $\Lambda_{\alpha_d} = \ln\vert\psi_{\alpha_d}\vert^2/2d -\ln(\Gamma/W)$ and $\Lambda_d=\max_{\alpha_d}\Lambda_{\alpha_d}$, the critical point $(\Gamma/W)_c$ can be expressed, using Eq.~\ref{eq:deloc-crit}, as 
\eq{
\begin{split}
&\lim_{d\to\infty}P(\Lambda_d>-\ln(\Gamma/W)_c)\to 1\\\Rightarrow & \lim_{d\to\infty}C(\Lambda_d=-\ln(\Gamma/W)_c)\to 1
\end{split}\,,
\label{eq:FSA-critical}
}
where $C(\Lambda_d)$ is the cumulative distribution and we have used the fact that the localisation length $\xi$ diverges at the transition. In other words, if $P(\Lambda_d)$ is peaked at some value $\braket{\Lambda_d}$ then the critical point is at $\ln(\Gamma/W)_c=-\braket{\Lambda_d}$.
Note that, as in the static case, taking the maximum amplitude over all sites at distance $d$ overestimates the critical disorder, resulting in an upper bound.
 
For the driven system at the $\Gamma=0$ limit, the $\ket{\varphi_a}$ are all localised on a single physical site and a single rung. 
Switching on $\Gamma$ allows for hopping to other sites and rungs. 
There is however one difference from the static case: for stroboscopic time evolution, $t=kT$ with integer $k$, $\ket{\phi_a(kT)}=e^{i\Omega_a kT}\sum_{n}\ket{\varphi_a^{(n)}}$. Thus, the amplitude at a physical site $\alpha$ will depend on the amplitudes on all rungs, $\sum_{n}\ket{\varphi_a^{(n)}}$, 
implying that the probability amplitude of the state spreading to physical site $\alpha_d$ is $\psi_{\alpha_d}=\sum_{n}\psi_{\alpha_d,n}$ where $\psi_{\alpha_d,n}$ is the amplitude on site $\alpha_d$ and Shirley rung $n$. 

To each path in the static system between two sites, say $\alpha_0\shortrightarrow\alpha_1\shortrightarrow\cdots\alpha_d$, there correspond $3^d$ paths in the driven system because at each hop in physical space the rung can change by $0$ or $\pm 1$ due to our model being monochromatically driven. Thus, these paths all originate at $\ket{\alpha,0}$ and terminate at $\ket{\alpha_d,n_d}$ with $n_d\in\left[-d,d\right]$~\footnote{The Shirley Hamiltonian \eqref{eq:H-shirley} is translation-invariant in the Shirley direction and the initial state`s Shirley index can be chosen to be $n=0$ without loss of generality}.

The total amplitude, $\psi_{\alpha_d}$ due to all these paths can then be written as
\eq{
\psi_{\alpha_d}=\sum_{p:\alpha_0\shortrightarrow\alpha_d}\sum_{q=1}^{3^d}\prod_{\alpha_i\in p}\frac{\Gamma}{\Delta_i+n_{\alpha_i}^{(q)}\omega}\,,
\label{eq:fsa-shirley}
}
where $\Delta_i=\epsilon_{\alpha_i}-\epsilon_{\alpha_0}$, the first sum (over $p$) is over the physical paths between $\alpha_0$ and $\alpha_d$, the second sum (over $q$) is over the Shirley paths corresponding to the physical path $p$, $\alpha_i$ is the physical site on the $i^\mathrm{th}$ step of the path, and $n_i^{(q)}$ is the rung index of the path $q$ at step $i$~\footnote{Formally, $n_i^{(q)}=\sum_{i=1}^q s_i^{(q)}$ where $s_i^{(q)}=\pm1,0$ is the change in the Shirley index on path $q$ at step $i$}.
The delocalisation criterion for the driven system is then identical to that in Eq.~\ref{eq:deloc-crit} but with the FSA amplitude obtained in the Shirley picture from Eq.~\ref{eq:fsa-shirley}. In the notation used in Eq.~\ref{eq:fsa-shirley}, the path present in the static case is the one for which $n_{\alpha_i}^{(q)}=0$ for all $i$.

It is important to distinguish between the various interference effects at play here.  For a given $n$, $\psi_{\alpha_d,n}$ receives contributions from multiple paths, each of which can be written as $\ket{\alpha_0,n_0}\rightarrow\ldots\rightarrow \ket{\alpha_j,n_j}\rightarrow\ldots\rightarrow\ket{\alpha_d,n_d}$ with $n_0=0$ and $n_d=n$. In particular, there are  (i) different physical paths that follow the same Shirley rungs, so the $\alpha_j$ are different for each path for $j\neq 0,d$ but the $n_j$ the same for all--this is present for the static case too; (ii) the same physical path on different rungs, so the $\alpha_j$ are the same for all but the $n_j$ different.
In addition, (iii)  since $\psi_{\alpha_d}=\sum_n\psi_{\alpha_d,n}$, the FSA amplitudes $\psi_{\alpha_d,n}$ on different rungs interfere with each other. 
Crucially, neither of (ii) and (iii) is present for the static case~\footnote{The role of interferences of this kind were studied in a different context of transport in a driven one-dimensional Anderson insulator~\cite{localisation2017agarwal}.}.
Eq.~\ref{eq:fsa-shirley} manifestly takes all of these kinds of interference into account.

In order to single out the effect of periodic-driving and interference between the Shirley paths, we take our physical graph to be a rooted tree, which has no loops.
There then exists a unique shortest physical path between any two sites of the graph, removing the interference effect labelled (i) above and leaving only the driving-induced effects (ii) and (iii).

For a tree with average branching number $K$~\footnote{$K$ can be fractional as well. For instance, if a fraction $p$ and $1-p$ of the sites have $K_1$ and $K_2$ descendants respectively, the average branching number is $K=pK_1 + (1-p)K_2$.}, there are $K^d$ sites at distance $d$ from the root and one needs the probability distribution of the maximum amplitude over these $K^d$ sites, see above Eq.~\ref{eq:deloc-crit}. As the amplitudes on each of these sites are independent and identically distributed with $P$, denoting this distribution by $\pmax{\Lambda_d}$, one finds
\eq{
	\pmax{\Lambda_d} = K^d [C(\Lambda_{d})]^{K^d-1}P(\Lambda_{d})\,,
	\label{eq:PLambda-K}
}
with $P,C$ the distribution and cumulative distribution for a single site.
Equivalently, $\cmax{\Lambda_d}=[C(\Lambda_{d})]^{K^d}$. It is thus possible to compute $\pmax{\Lambda_d}$ or $\cmax{\Lambda_d}$ for arbitrary $K$ by considering only single physical paths of length $d$. We, therefore, use the notation $\Lambda_d$ for both the amplitude of a single physical path and the maximum amplitude. The FSA for the former in the Shirley picture can be efficiently implemented using a transfer matrix~\cite{supp}.

In Fig.~\ref{fig:fsa} we show the numerical results from the FSA in the Shirley picture. The inset in panel (a) shows that the distributions $P(\Lambda_d)$ sharpen with increasing $d$. This is manifested in the cumulative distributions, $C(\Lambda_d)$ for various $d$ crossing at a particular value of $\Lambda_d$ as evidenced in panel (a).
The crossing point from Eq.~\ref{eq:FSA-critical} can be inferred to be $\ln(W/\Gamma)_c$, yielding the critical point. The critical $\Gamma_c/W$ so obtained is shown as a function of the frequency $\omega$ in panel (b). 
There are two crucial features of note: (i) For $\omega/W \gg 1$, $\Gamma_c/W$ is larger than that of the undriven system, $\Gamma_c^\mathrm{s}/W$. Thus, in this regime, the periodic driving \emph{enhances} the localised phase.
(ii) In the regime of $\omega/W \lessapprox 1$, $\Gamma_c/W$ decreases with decreasing $\omega$ and is, in fact, much smaller than $\Gamma_c^\mathrm{s}/W$. Hence, in this regime, the driving \emph{suppresses} the localised phase, favouring delocalisation, and parametrically so with decreasing $\omega$. 

In the following we present analytical arguments which give insight into the aforementioned behaviour of $\Gamma_c$ with $\omega$. As the distributions for a system with arbitrary $K$ are simply related to that for a single physical path via Eq.~\ref{eq:PLambda-K}, which leads to the qualitative behaviour of $\Gamma_c$ with $\omega$ being the same for all $K$, it suffices to analyse the FSA amplitudes for different Shirley paths corresponding to a single physical path $\alpha_0\shortrightarrow\alpha_1\shortrightarrow\cdots\alpha_d$. 
It will be useful to write  $\Lambda_d = \ln[(\sum_{q=1}^{3^d}w_q)^2]/2d$, where $(\Gamma/W)^d w_q$ is the amplitude of Shirley path $q$.
Further, since the distribution of $\Lambda_d$ gets sharper with increasing $d$, one may take $\lim_{d\to\infty}\braket{\Lambda_d}=-\ln(\Gamma/W)_c$.
It will also be useful to decompose $\Lambda_d$ into `direct' and `interference' terms,
\begin{equation}
	\Lambda_d = \underbrace{\frac{1}{2d}\ln\sum_q w_q^2}_{\Lambda_d^\mathrm{dir}}+\underbrace{\frac{1}{2d}\ln\left(1+\frac{\sum_{q\neq q^\prime}w_q w_{q^\prime}}{\sum_q w_q^2}\right)}_{\Lambda_d^\mathrm{int}}\,.
\end{equation}

We first turn to the low-frequency regime, $\omega/W\ll 1$.
We may drop the interference between the paths, approximating $\Lambda_d\approx\Lambda_d^\mathrm{dir}$ in this regime.
This is justified because~\cite{supp}, firstly, for $\omega/W\ll 1$, $\vert \Lambda_d^\mathrm{int}\vert\ll\vert \Lambda_d^\mathrm{dir}\vert$ and secondly $\Lambda_d^\mathrm{int}<0$ so that ignoring it overestimates $(W/\Gamma)_c$, consistent with the general bounds placed by the FSA.

\begin{figure}
    \includegraphics[width=\linewidth]{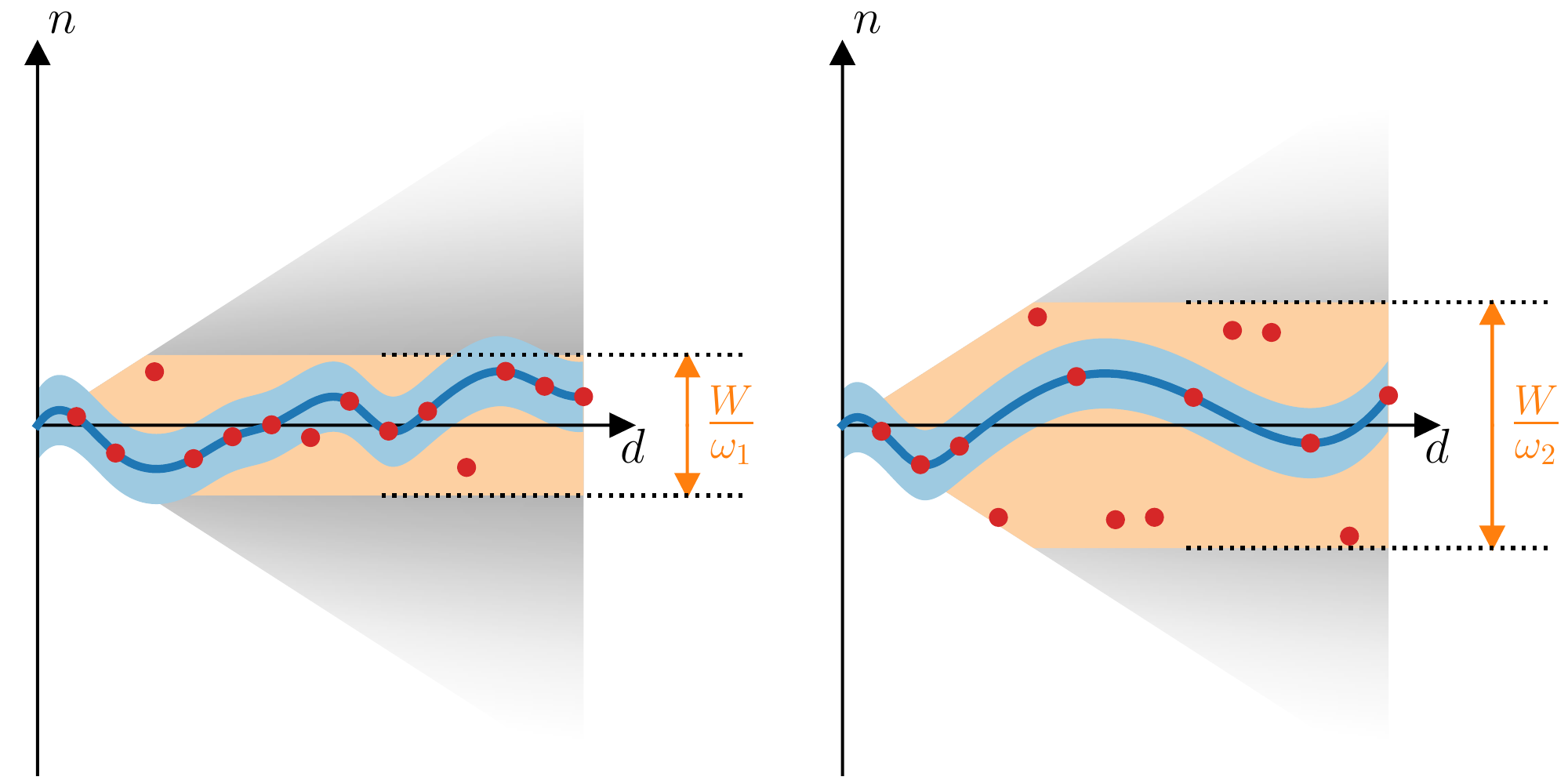}
    \caption{Schematic picture for the Shirley resonances in the low-frequency regime, $\omega\ll W$. In the plots above $\omega_1>\omega_2$. The resonances denoted by the red dots, where the effective site-energies $\vert \Delta_i+n_{\alpha_i}^\ast\omega\vert$ are within $\omega$ typically stay within a strip (shaded in orange) in the Shirley direction of width $W/\omega$. The larger $\omega/W$, the greater the likelihood of a resonance occurring per unit length. A path $q$ is resonant if the $n_\alpha^{(p)}$ passes within a distance 1 of these resonances, as indicated by the blue-shaded ``swept area'' of each path.}
    \label{fig:low-freq-schematic}
\end{figure}

For a given path $q$, ${\vert n_{\alpha_i}^{(q)}-n_{\alpha_{i+1}}^{(q)}\vert=0}$ or $1$, so a path selected at random from all possible paths is a simple random walk with $\alpha$ playing the role of time. On the other hand, the $\epsilon_\alpha$ are uncorrelated for different $\alpha$. 
We define a physical site $\alpha_i$ to be resonant on Shirley rung $n_{\alpha_i}^\ast$ if $\left|\Delta_i-n_{\alpha_i}^\ast\omega\right|<\omega$ such that all the resonances lie within a strip of width $\propto W/\omega$ in the Shirley direction (see Fig.~\ref{fig:low-freq-schematic}).
Hence, for a typical path, the probability of a resonance vanishes after $\gg(W/\omega)^2$ steps since by then the random walk is well outside the strip of resonances.
Although in the limit of $d\to\infty$, a fraction tending to 1 of these paths eventually do escape this strip eliminating the possibility of a resonance, driving can still cause delocalisation as $\Lambda_d$ consists of a sum of of the contributions of the paths and not an average.
One thus needs to consider an effective number of paths $N_q^\mathrm{eff}$ which is the sum, over all the paths, of the fraction of each path spent inside the strip.

For simplicity, consider a single path lying entirely within the strip. As it meanders along the strip (Fig.~\ref{fig:low-freq-schematic}) the probability it picks up a resonance per step is constant, $\rho(\omega)$, the same for all steps (due to the uncorrelated nature of the $\epsilon_\alpha$). A path of length $d$ will therefore pick up $\mu=\rho(\omega)d$ resonances on average. Over the ensemble of all paths lying entirely in the strip, the probability that a given path has $r$ resonances is Poisson, $P_r=\exp(-\mu)\mu^r/r!$. Finally, for a given path $q$, $w_q=\prod_{i=1}^d (\Delta_i+n^{(q)}_{\alpha_i})^{-1}$, we replace the factor for $\alpha_i$ with $1/W$ if $\alpha_i$ is off-resonant, $1/\omega$ if resonant; thus a path with $r$ resonances will have $w_q\approx (W/\omega)^{2r}$. 
From all the above together, we have $\Lambda_d\approx\frac{1}{2d}\ln\left(\npeff \sum_{r=0}^d P_r \frac{W^{2r}}{\omega^{2r}}\right)$. In the limit of $d\to\infty$,
\begin{equation}
	\lim_{d\to\infty}\Lambda_d\approx c-\frac{1}{2}\rho(\omega)\left(1-\frac{W^2}{\omega^2}\right)\,,
\end{equation}
where $c\equiv\lim_{d\to\infty}(\ln\npeff)/2d$ is an $\mathcal{O}(1)$ constant~\cite{supp}.
As a site $\alpha_i$ on some path $q$ is resonant when $\vert\Delta_i+n_{\alpha_i}^{(q)}\vert$ is within $\omega$, a rough estimate for the probability of a resonance $\rho(\omega)$ is $\omega/W$. With this estimate, the localisation critical point is
\begin{equation}
	(\Gamma/W)_c\approx a\,\exp(-bW/\omega)\,,
	\label{eq:Gammac-low-freq}	
\end{equation}
where $a$ and $b$ are constants. Thus, in this regime, driving destabilises the localised phase (suppressing $\Gamma_c$) as one might expect~\cite{lazarides2015fate,ponte2015many,abanin2016theory}. Note that this is a result of the competition between the increase in the number of resonances (since $\rho(\omega)=\omega/W$) and the reduction of their strength ($1/\omega$) as $\omega$ increases.

Next we turn to the high-frequency regime, $\omega/W\gg 1$, where the driving enhances localisation as indicated by the increase in $\Gamma_c$ compared to $\Gamma_c^\mathrm{s}$.
We first note that for $\omega=\infty$, only the static path contributes, the amplitude for which we denote as $w_\mathrm{s}$. We will therefore use it as a reference and consider the effect of deviations from it.
For large $\omega/W\gg 1$, Shirley paths that deviate from the static path at $x$ sites have an amplitude relative to the static one suppressed parametrically as $\mathcal{O}((W/\omega)^x)$. 
The leading order contribution (at $\mathcal{O}((W/\omega)^{2})$ ) can be shown to be~\cite{supp}
\eq{
\Lambda_d(\omega) \approx \frac{1}{2d}\ln\left[w_\mathrm{s}^2\left(1-\frac{4}{\omega^2}\sum_{i=1}^d\Delta_i^2+\frac{\Delta_{d-1}\Delta_{d}}{\omega^2}\right)\right]\,.
\label{eq:Lambda-d-highfreq}
}
Equation~\ref{eq:Lambda-d-highfreq} already shows that $\Lambda_d$ in the presence of driving is smaller than that of the static case, $\Lambda_d^\mathrm{s}=\ln w_\mathrm{s}^2/2d$, for any disorder realisation indicating an enhancement of the localised phase.
Using the fact that the $\Delta_i$'s are independent of each other and $\omega/\Delta_i\gg1$, Eq.~\ref{eq:Lambda-d-highfreq} gives
\eq{
\braket{\Lambda_d(\omega)}\overset{d\gg1}{\approx}\braket{\Lambda_d^\mathrm{s}}-4W^2/\omega^2\,.
\label{eq:Lambda-d-high-freq-shift}
}
Since the distribution of $\Lambda_d$ sharpens with increasing $d$ (see Fig.~\ref{fig:fsa}(a)), it seems reasonable to assume that $\lim_{d\to\infty}\braket{\Lambda_d}=-\ln(\Gamma/W)_c$, as such Eq.~\ref{eq:Lambda-d-high-freq-shift} yields,
\eq{
\Gamma_c(\omega) \approx \Gamma_c^\mathrm{s}e^{4W^2/\omega^2}\,.
\label{eq:Gammac-high-freq}
}
The result in Eq.~\ref{eq:Gammac-high-freq} explicitly shows the increase in $\Gamma_c$ from $\Gamma_c^\mathrm{s}$ and the enhancement of the localised phase.

The mechanism behind the driving-induced suppression of the total FSA amplitude can be understood as follows. 
For every site $\alpha_i$ on the physical path, there exists a pair of Shirley paths with $n_{\alpha_j}^{(q)}=\pm \delta_{ij}$. 
If we consider, without loss generality, $\Delta_i>0$, then $\Delta_i\pm \omega \gtrless 0$ with $\vert\Delta_i+ \omega\vert>\vert\Delta_i-\omega\vert$.
Hence, the path whose amplitude has the opposite (same) sign to the static path has a higher (lower) magnitude.
The total amplitude is thus suppressed compared to the static case. However, the correction turns out to be $\mathcal{O}(1/\omega^2)$ so that one has to take into account all paths which deviate at up to two sites from the static path.

The physical picture and the derivation of Eq.~\ref{eq:Lambda-d-highfreq} suggest that this generalises to pairs of paths that deviate at multiple sites from the static path, but with opposite signs of the Shirley rung indices.
While the quantitative result in Eq.~\ref{eq:Gammac-high-freq} can change upon including higher-order (in $1/\omega$) contributions, localisation enhancement is expected to stay robust; however, a concrete analytical demonstration remains  for future work.

In summary, we have addressed the problem of Anderson localisation on driven, disordered trees using the FSA on an extended Shirley graph by eliminating the time-dependence of the Hamiltonian in favour of an extra dimension.
The interplay between the availability of additional paths due to the driving and their interference yields a non-trivial localisation phase diagram; in the high-frequency regime, the localised phase is stabilised relative to the undriven case, so that there is localisation for weaker disorder than in the static case, whereas in the low-frequency regime driving destabilises localisation and increases the strength of the disorder required for localisation.

A natural next step towards addressing the stability towards driving of many-body localisation due to either correlations in Fock-space disorder~\cite{roy2020fock} or constraints in the Fock-space graph~\cite{roy2020strong} would be to incorporate the effect of multiple static paths between any two sites, rather than focus on purely tree-like structures as we have done here.
It is also conceivable that the framework discussed here can be used to analyse the problem of localisation or lack thereof in quasi-periodically driven systems where one would require multiple Shirley directions, one for each of the frequencies present in the drive.

\begin{acknowledgments}
SR thanks David E. Logan, and RM thanks Antonello Scardicchio and Shivaji Sondhi for discussions of the FSA.
This work was in part supported by EPSRC Grant No. EP/S020527/1 and by the Deutsche Forschungsgemeinschaft  under  cluster of excellence ct.qmat (EXC 2147, project-id 390858490).
\end{acknowledgments}

\bibliography{refs}

\begin{thebibliography}{47}%
\makeatletter
\providecommand \@ifxundefined [1]{%
 \@ifx{#1\undefined}
}%
\providecommand \@ifnum [1]{%
 \ifnum #1\expandafter \@firstoftwo
 \else \expandafter \@secondoftwo
 \fi
}%
\providecommand \@ifx [1]{%
 \ifx #1\expandafter \@firstoftwo
 \else \expandafter \@secondoftwo
 \fi
}%
\providecommand \natexlab [1]{#1}%
\providecommand \enquote  [1]{``#1''}%
\providecommand \bibnamefont  [1]{#1}%
\providecommand \bibfnamefont [1]{#1}%
\providecommand \citenamefont [1]{#1}%
\providecommand \href@noop [0]{\@secondoftwo}%
\providecommand \href [0]{\begingroup \@sanitize@url \@href}%
\providecommand \@href[1]{\@@startlink{#1}\@@href}%
\providecommand \@@href[1]{\endgroup#1\@@endlink}%
\providecommand \@sanitize@url [0]{\catcode `\\12\catcode `\$12\catcode
  `\&12\catcode `\#12\catcode `\^12\catcode `\_12\catcode `\%12\relax}%
\providecommand \@@startlink[1]{}%
\providecommand \@@endlink[0]{}%
\providecommand \url  [0]{\begingroup\@sanitize@url \@url }%
\providecommand \@url [1]{\endgroup\@href {#1}{\urlprefix }}%
\providecommand \urlprefix  [0]{URL }%
\providecommand \Eprint [0]{\href }%
\providecommand \doibase [0]{https://doi.org/}%
\providecommand \selectlanguage [0]{\@gobble}%
\providecommand \bibinfo  [0]{\@secondoftwo}%
\providecommand \bibfield  [0]{\@secondoftwo}%
\providecommand \translation [1]{[#1]}%
\providecommand \BibitemOpen [0]{}%
\providecommand \bibitemStop [0]{}%
\providecommand \bibitemNoStop [0]{.\EOS\space}%
\providecommand \EOS [0]{\spacefactor3000\relax}%
\providecommand \BibitemShut  [1]{\csname bibitem#1\endcsname}%
\let\auto@bib@innerbib\@empty
\bibitem [{\citenamefont {Anderson}(1958)}]{anderson1958absence}%
  \BibitemOpen
  \bibfield  {author} {\bibinfo {author} {\bibfnamefont {P.~W.}\ \bibnamefont
  {Anderson}},\ }\bibfield  {title} {\bibinfo {title} {Absence of diffusion in
  certain random lattices},\ }\href {https://doi.org/10.1103/PhysRev.109.1492}
  {\bibfield  {journal} {\bibinfo  {journal} {Phys. Rev.}\ }\textbf {\bibinfo
  {volume} {109}},\ \bibinfo {pages} {1492} (\bibinfo {year}
  {1958})}\BibitemShut {NoStop}%
\bibitem [{\citenamefont {Basko}\ \emph {et~al.}(2006)\citenamefont {Basko},
  \citenamefont {Aleiner},\ and\ \citenamefont {Altshuler}}]{basko2006metal}%
  \BibitemOpen
  \bibfield  {author} {\bibinfo {author} {\bibfnamefont {D.~M.}\ \bibnamefont
  {Basko}}, \bibinfo {author} {\bibfnamefont {I.~L.}\ \bibnamefont {Aleiner}},\
  and\ \bibinfo {author} {\bibfnamefont {B.~L.}\ \bibnamefont {Altshuler}},\
  }\bibfield  {title} {\bibinfo {title} {Metal--insulator transition in a
  weakly interacting many-electron system with localized single-particle
  states},\ }\href
  {http://www.sciencedirect.com/science/article/pii/S0003491605002630}
  {\bibfield  {journal} {\bibinfo  {journal} {Annals of {P}hysics}\ }\textbf
  {\bibinfo {volume} {321}},\ \bibinfo {pages} {1126} (\bibinfo {year}
  {2006})}\BibitemShut {NoStop}%
\bibitem [{\citenamefont {Gornyi}\ \emph {et~al.}(2005)\citenamefont {Gornyi},
  \citenamefont {Mirlin},\ and\ \citenamefont
  {Polyakov}}]{gornyi2005interacting}%
  \BibitemOpen
  \bibfield  {author} {\bibinfo {author} {\bibfnamefont {I.~V.}\ \bibnamefont
  {Gornyi}}, \bibinfo {author} {\bibfnamefont {A.~D.}\ \bibnamefont {Mirlin}},\
  and\ \bibinfo {author} {\bibfnamefont {D.~G.}\ \bibnamefont {Polyakov}},\
  }\bibfield  {title} {\bibinfo {title} {Interacting electrons in disordered
  wires: Anderson localization and low-${T}$ transport},\ }\href
  {https://doi.org/10.1103/PhysRevLett.95.206603} {\bibfield  {journal}
  {\bibinfo  {journal} {Phys. Rev. Lett.}\ }\textbf {\bibinfo {volume} {95}},\
  \bibinfo {pages} {206603} (\bibinfo {year} {2005})}\BibitemShut {NoStop}%
\bibitem [{\citenamefont {Oganesyan}\ and\ \citenamefont
  {Huse}(2007)}]{oganesyan2007localisation}%
  \BibitemOpen
  \bibfield  {author} {\bibinfo {author} {\bibfnamefont {V.}~\bibnamefont
  {Oganesyan}}\ and\ \bibinfo {author} {\bibfnamefont {D.~A.}\ \bibnamefont
  {Huse}},\ }\bibfield  {title} {\bibinfo {title} {Localization of interacting
  fermions at high temperature},\ }\href
  {https://doi.org/10.1103/PhysRevB.75.155111} {\bibfield  {journal} {\bibinfo
  {journal} {Phys. Rev. B}\ }\textbf {\bibinfo {volume} {75}},\ \bibinfo
  {pages} {155111} (\bibinfo {year} {2007})}\BibitemShut {NoStop}%
\bibitem [{\citenamefont {\ifmmode \check{Z}\else
  \v{Z}\fi{}nidari\ifmmode~\check{c}\else \v{c}\fi{}}\ \emph
  {et~al.}(2008)\citenamefont {\ifmmode \check{Z}\else
  \v{Z}\fi{}nidari\ifmmode~\check{c}\else \v{c}\fi{}}, \citenamefont {Prosen},\
  and\ \citenamefont {Prelov\ifmmode~\check{s}\else
  \v{s}\fi{}ek}}]{znidaric2008many}%
  \BibitemOpen
  \bibfield  {author} {\bibinfo {author} {\bibfnamefont {M.}~\bibnamefont
  {\ifmmode \check{Z}\else \v{Z}\fi{}nidari\ifmmode~\check{c}\else
  \v{c}\fi{}}}, \bibinfo {author} {\bibfnamefont {T.}~\bibnamefont {Prosen}},\
  and\ \bibinfo {author} {\bibfnamefont {P.}~\bibnamefont
  {Prelov\ifmmode~\check{s}\else \v{s}\fi{}ek}},\ }\bibfield  {title} {\bibinfo
  {title} {Many-body localization in the {H}eisenberg {XXZ} magnet in a random
  field},\ }\href {https://doi.org/10.1103/PhysRevB.77.064426} {\bibfield
  {journal} {\bibinfo  {journal} {Phys. Rev. B}\ }\textbf {\bibinfo {volume}
  {77}},\ \bibinfo {pages} {064426} (\bibinfo {year} {2008})}\BibitemShut
  {NoStop}%
\bibitem [{\citenamefont {Nandkishore}\ and\ \citenamefont
  {Huse}(2015)}]{nandkishore2015many}%
  \BibitemOpen
  \bibfield  {author} {\bibinfo {author} {\bibfnamefont {R.}~\bibnamefont
  {Nandkishore}}\ and\ \bibinfo {author} {\bibfnamefont {D.~A.}\ \bibnamefont
  {Huse}},\ }\bibfield  {title} {\bibinfo {title} {Many-body localization and
  thermalization in quantum statistical mechanics},\ }\href
  {https://doi.org/10.1146/annurev-conmatphys-031214-014726} {\bibfield
  {journal} {\bibinfo  {journal} {Annu. Rev. Condens. Matter Phys.}\ }\textbf
  {\bibinfo {volume} {6}},\ \bibinfo {pages} {15} (\bibinfo {year}
  {2015})}\BibitemShut {NoStop}%
\bibitem [{\citenamefont {Huse}\ \emph {et~al.}(2013)\citenamefont {Huse},
  \citenamefont {Nandkishore}, \citenamefont {Oganesyan}, \citenamefont {Pal},\
  and\ \citenamefont {Sondhi}}]{huse2013localisation}%
  \BibitemOpen
  \bibfield  {author} {\bibinfo {author} {\bibfnamefont {D.~A.}\ \bibnamefont
  {Huse}}, \bibinfo {author} {\bibfnamefont {R.}~\bibnamefont {Nandkishore}},
  \bibinfo {author} {\bibfnamefont {V.}~\bibnamefont {Oganesyan}}, \bibinfo
  {author} {\bibfnamefont {A.}~\bibnamefont {Pal}},\ and\ \bibinfo {author}
  {\bibfnamefont {S.~L.}\ \bibnamefont {Sondhi}},\ }\bibfield  {title}
  {\bibinfo {title} {Localization-protected quantum order},\ }\href
  {https://doi.org/10.1103/PhysRevB.88.014206} {\bibfield  {journal} {\bibinfo
  {journal} {Phys. Rev. B}\ }\textbf {\bibinfo {volume} {88}},\ \bibinfo
  {pages} {014206} (\bibinfo {year} {2013})}\BibitemShut {NoStop}%
\bibitem [{\citenamefont {Khemani}\ \emph {et~al.}(2016)\citenamefont
  {Khemani}, \citenamefont {Lazarides}, \citenamefont {Moessner},\ and\
  \citenamefont {Sondhi}}]{khemani2016phase}%
  \BibitemOpen
  \bibfield  {author} {\bibinfo {author} {\bibfnamefont {V.}~\bibnamefont
  {Khemani}}, \bibinfo {author} {\bibfnamefont {A.}~\bibnamefont {Lazarides}},
  \bibinfo {author} {\bibfnamefont {R.}~\bibnamefont {Moessner}},\ and\
  \bibinfo {author} {\bibfnamefont {S.~L.}\ \bibnamefont {Sondhi}},\ }\bibfield
   {title} {\bibinfo {title} {Phase structure of driven quantum systems},\
  }\href {https://doi.org/10.1103/PhysRevLett.116.250401} {\bibfield  {journal}
  {\bibinfo  {journal} {Phys. Rev. Lett.}\ }\textbf {\bibinfo {volume} {116}},\
  \bibinfo {pages} {250401} (\bibinfo {year} {2016})}\BibitemShut {NoStop}%
\bibitem [{\citenamefont {Else}\ \emph {et~al.}(2016)\citenamefont {Else},
  \citenamefont {Bauer},\ and\ \citenamefont {Nayak}}]{else2016floquet}%
  \BibitemOpen
  \bibfield  {author} {\bibinfo {author} {\bibfnamefont {D.~V.}\ \bibnamefont
  {Else}}, \bibinfo {author} {\bibfnamefont {B.}~\bibnamefont {Bauer}},\ and\
  \bibinfo {author} {\bibfnamefont {C.}~\bibnamefont {Nayak}},\ }\bibfield
  {title} {\bibinfo {title} {Floquet time crystals},\ }\href
  {https://doi.org/10.1103/PhysRevLett.117.090402} {\bibfield  {journal}
  {\bibinfo  {journal} {Phys. Rev. Lett.}\ }\textbf {\bibinfo {volume} {117}},\
  \bibinfo {pages} {090402} (\bibinfo {year} {2016})}\BibitemShut {NoStop}%
\bibitem [{\citenamefont {Yao}\ \emph {et~al.}(2017)\citenamefont {Yao},
  \citenamefont {Potter}, \citenamefont {Potirniche},\ and\ \citenamefont
  {Vishwanath}}]{yao2017discrete}%
  \BibitemOpen
  \bibfield  {author} {\bibinfo {author} {\bibfnamefont {N.~Y.}\ \bibnamefont
  {Yao}}, \bibinfo {author} {\bibfnamefont {A.~C.}\ \bibnamefont {Potter}},
  \bibinfo {author} {\bibfnamefont {I.-D.}\ \bibnamefont {Potirniche}},\ and\
  \bibinfo {author} {\bibfnamefont {A.}~\bibnamefont {Vishwanath}},\ }\bibfield
   {title} {\bibinfo {title} {Discrete time crystals: Rigidity, criticality,
  and realizations},\ }\href {https://doi.org/10.1103/PhysRevLett.118.030401}
  {\bibfield  {journal} {\bibinfo  {journal} {Phys. Rev. Lett.}\ }\textbf
  {\bibinfo {volume} {118}},\ \bibinfo {pages} {030401} (\bibinfo {year}
  {2017})}\BibitemShut {NoStop}%
\bibitem [{\citenamefont {Moessner}\ and\ \citenamefont
  {Sondhi}(2017)}]{moessner2017equilibration}%
  \BibitemOpen
  \bibfield  {author} {\bibinfo {author} {\bibfnamefont {R.}~\bibnamefont
  {Moessner}}\ and\ \bibinfo {author} {\bibfnamefont {S.~L.}\ \bibnamefont
  {Sondhi}},\ }\bibfield  {title} {\bibinfo {title} {Equilibration and order in
  quantum {F}loquet matter},\ }\href
  {https://www.nature.com/nphys/journal/v13/n5/abs/nphys4106.html} {\bibfield
  {journal} {\bibinfo  {journal} {Nat. Phys.}\ }\textbf {\bibinfo {volume}
  {13}},\ \bibinfo {pages} {424} (\bibinfo {year} {2017})}\BibitemShut
  {NoStop}%
\bibitem [{\citenamefont {Watanabe}\ and\ \citenamefont
  {Oshikawa}(2015)}]{watanabe2015absence}%
  \BibitemOpen
  \bibfield  {author} {\bibinfo {author} {\bibfnamefont {H.}~\bibnamefont
  {Watanabe}}\ and\ \bibinfo {author} {\bibfnamefont {M.}~\bibnamefont
  {Oshikawa}},\ }\bibfield  {title} {\bibinfo {title} {Absence of quantum time
  crystals},\ }\href {https://doi.org/10.1103/PhysRevLett.114.251603}
  {\bibfield  {journal} {\bibinfo  {journal} {Phys. Rev. Lett.}\ }\textbf
  {\bibinfo {volume} {114}},\ \bibinfo {pages} {251603} (\bibinfo {year}
  {2015})}\BibitemShut {NoStop}%
\bibitem [{\citenamefont {Khemani}\ \emph {et~al.}(2019)\citenamefont
  {Khemani}, \citenamefont {Moessner},\ and\ \citenamefont
  {Sondhi}}]{khemani2019brief}%
  \BibitemOpen
  \bibfield  {author} {\bibinfo {author} {\bibfnamefont {V.}~\bibnamefont
  {Khemani}}, \bibinfo {author} {\bibfnamefont {R.}~\bibnamefont {Moessner}},\
  and\ \bibinfo {author} {\bibfnamefont {S.~L.}\ \bibnamefont {Sondhi}},\
  }\href@noop {} {\bibinfo {title} {A {B}rief {H}istory of {T}ime {C}rystals}}
  (\bibinfo {year} {2019}),\ \Eprint {https://arxiv.org/abs/1910.10745}
  {arXiv:1910.10745 [cond-mat.str-el]} \BibitemShut {NoStop}%
\bibitem [{\citenamefont {Lazarides}\ \emph {et~al.}(2014)\citenamefont
  {Lazarides}, \citenamefont {Das},\ and\ \citenamefont
  {Moessner}}]{lazarides2014equilibrium}%
  \BibitemOpen
  \bibfield  {author} {\bibinfo {author} {\bibfnamefont {A.}~\bibnamefont
  {Lazarides}}, \bibinfo {author} {\bibfnamefont {A.}~\bibnamefont {Das}},\
  and\ \bibinfo {author} {\bibfnamefont {R.}~\bibnamefont {Moessner}},\
  }\bibfield  {title} {\bibinfo {title} {Equilibrium states of generic quantum
  systems subject to periodic driving},\ }\href
  {https://doi.org/10.1103/PhysRevE.90.012110} {\bibfield  {journal} {\bibinfo
  {journal} {Phys. Rev. E}\ }\textbf {\bibinfo {volume} {90}},\ \bibinfo
  {pages} {012110} (\bibinfo {year} {2014})}\BibitemShut {NoStop}%
\bibitem [{\citenamefont {D'Alessio}\ and\ \citenamefont
  {Rigol}(2014)}]{dalessio2014long}%
  \BibitemOpen
  \bibfield  {author} {\bibinfo {author} {\bibfnamefont {L.}~\bibnamefont
  {D'Alessio}}\ and\ \bibinfo {author} {\bibfnamefont {M.}~\bibnamefont
  {Rigol}},\ }\bibfield  {title} {\bibinfo {title} {Long-time behavior of
  isolated periodically driven interacting lattice systems},\ }\href
  {https://doi.org/10.1103/PhysRevX.4.041048} {\bibfield  {journal} {\bibinfo
  {journal} {Phys. Rev. X}\ }\textbf {\bibinfo {volume} {4}},\ \bibinfo {pages}
  {041048} (\bibinfo {year} {2014})}\BibitemShut {NoStop}%
\bibitem [{\citenamefont {Ponte}\ \emph
  {et~al.}(2015{\natexlab{a}})\citenamefont {Ponte}, \citenamefont
  {Papi\ifmmode~\acute{c}\else \'{c}\fi{}}, \citenamefont {Huveneers},\ and\
  \citenamefont {Abanin}}]{ponte2015many}%
  \BibitemOpen
  \bibfield  {author} {\bibinfo {author} {\bibfnamefont {P.}~\bibnamefont
  {Ponte}}, \bibinfo {author} {\bibfnamefont {Z.}~\bibnamefont
  {Papi\ifmmode~\acute{c}\else \'{c}\fi{}}}, \bibinfo {author} {\bibfnamefont
  {F.}~\bibnamefont {Huveneers}},\ and\ \bibinfo {author} {\bibfnamefont
  {D.~A.}\ \bibnamefont {Abanin}},\ }\bibfield  {title} {\bibinfo {title}
  {Many-body localization in periodically driven systems},\ }\href
  {https://doi.org/10.1103/PhysRevLett.114.140401} {\bibfield  {journal}
  {\bibinfo  {journal} {Phys. Rev. Lett.}\ }\textbf {\bibinfo {volume} {114}},\
  \bibinfo {pages} {140401} (\bibinfo {year} {2015}{\natexlab{a}})}\BibitemShut
  {NoStop}%
\bibitem [{\citenamefont {Lazarides}\ \emph {et~al.}(2015)\citenamefont
  {Lazarides}, \citenamefont {Das},\ and\ \citenamefont
  {Moessner}}]{lazarides2015fate}%
  \BibitemOpen
  \bibfield  {author} {\bibinfo {author} {\bibfnamefont {A.}~\bibnamefont
  {Lazarides}}, \bibinfo {author} {\bibfnamefont {A.}~\bibnamefont {Das}},\
  and\ \bibinfo {author} {\bibfnamefont {R.}~\bibnamefont {Moessner}},\
  }\bibfield  {title} {\bibinfo {title} {Fate of many-body localization under
  periodic driving},\ }\href {https://doi.org/10.1103/PhysRevLett.115.030402}
  {\bibfield  {journal} {\bibinfo  {journal} {Phys. Rev. Lett.}\ }\textbf
  {\bibinfo {volume} {115}},\ \bibinfo {pages} {030402} (\bibinfo {year}
  {2015})}\BibitemShut {NoStop}%
\bibitem [{\citenamefont {Ponte}\ \emph
  {et~al.}(2015{\natexlab{b}})\citenamefont {Ponte}, \citenamefont {Chandran},
  \citenamefont {Papi{\'c}},\ and\ \citenamefont
  {Abanin}}]{ponte2015periodically}%
  \BibitemOpen
  \bibfield  {author} {\bibinfo {author} {\bibfnamefont {P.}~\bibnamefont
  {Ponte}}, \bibinfo {author} {\bibfnamefont {A.}~\bibnamefont {Chandran}},
  \bibinfo {author} {\bibfnamefont {Z.}~\bibnamefont {Papi{\'c}}},\ and\
  \bibinfo {author} {\bibfnamefont {D.~A.}\ \bibnamefont {Abanin}},\ }\bibfield
   {title} {\bibinfo {title} {Periodically driven ergodic and many-body
  localized quantum systems},\ }\href
  {https://www.sciencedirect.com/science/article/pii/S0003491614003212}
  {\bibfield  {journal} {\bibinfo  {journal} {Annals of Physics}\ }\textbf
  {\bibinfo {volume} {353}},\ \bibinfo {pages} {196} (\bibinfo {year}
  {2015}{\natexlab{b}})}\BibitemShut {NoStop}%
\bibitem [{\citenamefont {Abanin}\ \emph {et~al.}(2016)\citenamefont {Abanin},
  \citenamefont {{De Roeck}},\ and\ \citenamefont
  {Huveneers}}]{abanin2016theory}%
  \BibitemOpen
  \bibfield  {author} {\bibinfo {author} {\bibfnamefont {D.~A.}\ \bibnamefont
  {Abanin}}, \bibinfo {author} {\bibfnamefont {W.}~\bibnamefont {{De Roeck}}},\
  and\ \bibinfo {author} {\bibfnamefont {F.}~\bibnamefont {Huveneers}},\
  }\bibfield  {title} {\bibinfo {title} {Theory of many-body localization in
  periodically driven systems},\ }\href
  {https://doi.org/https://doi.org/10.1016/j.aop.2016.03.010} {\bibfield
  {journal} {\bibinfo  {journal} {Annals of Physics}\ }\textbf {\bibinfo
  {volume} {372}},\ \bibinfo {pages} {1 } (\bibinfo {year} {2016})}\BibitemShut
  {NoStop}%
\bibitem [{\citenamefont {Roy}\ and\ \citenamefont
  {Logan}(2020{\natexlab{a}})}]{roy2020fock}%
  \BibitemOpen
  \bibfield  {author} {\bibinfo {author} {\bibfnamefont {S.}~\bibnamefont
  {Roy}}\ and\ \bibinfo {author} {\bibfnamefont {D.~E.}\ \bibnamefont
  {Logan}},\ }\bibfield  {title} {\bibinfo {title} {Fock-space correlations and
  the origins of many-body localization},\ }\href
  {https://doi.org/10.1103/PhysRevB.101.134202} {\bibfield  {journal} {\bibinfo
   {journal} {Phys. Rev. B}\ }\textbf {\bibinfo {volume} {101}},\ \bibinfo
  {pages} {134202} (\bibinfo {year} {2020}{\natexlab{a}})}\BibitemShut
  {NoStop}%
\bibitem [{\citenamefont {Roy}\ and\ \citenamefont
  {Logan}(2020{\natexlab{b}})}]{roy2020localisation}%
  \BibitemOpen
  \bibfield  {author} {\bibinfo {author} {\bibfnamefont {S.}~\bibnamefont
  {Roy}}\ and\ \bibinfo {author} {\bibfnamefont {D.~E.}\ \bibnamefont
  {Logan}},\ }\bibfield  {title} {\bibinfo {title} {Localization on certain
  graphs with strongly correlated disorder},\ }\href
  {https://doi.org/10.1103/PhysRevLett.125.250402} {\bibfield  {journal}
  {\bibinfo  {journal} {Phys. Rev. Lett.}\ }\textbf {\bibinfo {volume} {125}},\
  \bibinfo {pages} {250402} (\bibinfo {year} {2020}{\natexlab{b}})}\BibitemShut
  {NoStop}%
\bibitem [{\citenamefont {Abou-Chacra}\ \emph {et~al.}(1973)\citenamefont
  {Abou-Chacra}, \citenamefont {Thouless},\ and\ \citenamefont
  {Anderson}}]{abou-chacra1973self}%
  \BibitemOpen
  \bibfield  {author} {\bibinfo {author} {\bibfnamefont {R.}~\bibnamefont
  {Abou-Chacra}}, \bibinfo {author} {\bibfnamefont {D.~J.}\ \bibnamefont
  {Thouless}},\ and\ \bibinfo {author} {\bibfnamefont {P.~W.}\ \bibnamefont
  {Anderson}},\ }\bibfield  {title} {\bibinfo {title} {A self-consistent theory
  of localization},\ }\href {https://doi.org/10.1088/0022-3719/6/10/009}
  {\bibfield  {journal} {\bibinfo  {journal} {Journal of Physics C: Solid State
  Physics}\ }\textbf {\bibinfo {volume} {6}},\ \bibinfo {pages} {1734}
  (\bibinfo {year} {1973})}\BibitemShut {NoStop}%
\bibitem [{\citenamefont {Chalker}\ and\ \citenamefont
  {Siak}(1990)}]{chalker1990anderson}%
  \BibitemOpen
  \bibfield  {author} {\bibinfo {author} {\bibfnamefont {J.~T.}\ \bibnamefont
  {Chalker}}\ and\ \bibinfo {author} {\bibfnamefont {S.}~\bibnamefont {Siak}},\
  }\bibfield  {title} {\bibinfo {title} {Anderson localisation on a {C}ayley
  tree: a new model with a simple solution},\ }\href
  {https://doi.org/10.1088/0953-8984/2/11/011} {\bibfield  {journal} {\bibinfo
  {journal} {J. Phys.: Cond. Matt.}\ }\textbf {\bibinfo {volume} {2}},\
  \bibinfo {pages} {2671} (\bibinfo {year} {1990})}\BibitemShut {NoStop}%
\bibitem [{\citenamefont {Evers}\ and\ \citenamefont
  {Mirlin}(2008)}]{evers2008anderson}%
  \BibitemOpen
  \bibfield  {author} {\bibinfo {author} {\bibfnamefont {F.}~\bibnamefont
  {Evers}}\ and\ \bibinfo {author} {\bibfnamefont {A.~D.}\ \bibnamefont
  {Mirlin}},\ }\bibfield  {title} {\bibinfo {title} {Anderson transitions},\
  }\href {https://doi.org/10.1103/RevModPhys.80.1355} {\bibfield  {journal}
  {\bibinfo  {journal} {Rev. Mod. Phys.}\ }\textbf {\bibinfo {volume} {80}},\
  \bibinfo {pages} {1355} (\bibinfo {year} {2008})}\BibitemShut {NoStop}%
\bibitem [{\citenamefont {De~Luca}\ \emph {et~al.}(2014)\citenamefont
  {De~Luca}, \citenamefont {Altshuler}, \citenamefont {Kravtsov},\ and\
  \citenamefont {Scardicchio}}]{luca2014anderson}%
  \BibitemOpen
  \bibfield  {author} {\bibinfo {author} {\bibfnamefont {A.}~\bibnamefont
  {De~Luca}}, \bibinfo {author} {\bibfnamefont {B.~L.}\ \bibnamefont
  {Altshuler}}, \bibinfo {author} {\bibfnamefont {V.~E.}\ \bibnamefont
  {Kravtsov}},\ and\ \bibinfo {author} {\bibfnamefont {A.}~\bibnamefont
  {Scardicchio}},\ }\bibfield  {title} {\bibinfo {title} {Anderson localization
  on the {B}ethe lattice: Nonergodicity of extended states},\ }\href
  {https://doi.org/10.1103/PhysRevLett.113.046806} {\bibfield  {journal}
  {\bibinfo  {journal} {Phys. Rev. Lett.}\ }\textbf {\bibinfo {volume} {113}},\
  \bibinfo {pages} {046806} (\bibinfo {year} {2014})}\BibitemShut {NoStop}%
\bibitem [{\citenamefont {Altshuler}\ \emph {et~al.}(2016)\citenamefont
  {Altshuler}, \citenamefont {Ioffe},\ and\ \citenamefont
  {Kravtsov}}]{altshuler2016multifractal}%
  \BibitemOpen
  \bibfield  {author} {\bibinfo {author} {\bibfnamefont {B.~L.}\ \bibnamefont
  {Altshuler}}, \bibinfo {author} {\bibfnamefont {L.~B.}\ \bibnamefont
  {Ioffe}},\ and\ \bibinfo {author} {\bibfnamefont {V.~E.}\ \bibnamefont
  {Kravtsov}},\ }\href@noop {} {\bibinfo {title} {Multifractal states in
  self-consistent theory of localization: analytical solution}} (\bibinfo
  {year} {2016}),\ \Eprint {https://arxiv.org/abs/1610.00758} {arXiv:1610.00758
  [cond-mat.dis-nn]} \BibitemShut {NoStop}%
\bibitem [{\citenamefont {Tikhonov}\ \emph {et~al.}(2016)\citenamefont
  {Tikhonov}, \citenamefont {Mirlin},\ and\ \citenamefont
  {Skvortsov}}]{tikhonov2016anderson}%
  \BibitemOpen
  \bibfield  {author} {\bibinfo {author} {\bibfnamefont {K.~S.}\ \bibnamefont
  {Tikhonov}}, \bibinfo {author} {\bibfnamefont {A.~D.}\ \bibnamefont
  {Mirlin}},\ and\ \bibinfo {author} {\bibfnamefont {M.~A.}\ \bibnamefont
  {Skvortsov}},\ }\bibfield  {title} {\bibinfo {title} {Anderson localization
  and ergodicity on random regular graphs},\ }\href
  {https://doi.org/10.1103/PhysRevB.94.220203} {\bibfield  {journal} {\bibinfo
  {journal} {Phys. Rev. B}\ }\textbf {\bibinfo {volume} {94}},\ \bibinfo
  {pages} {220203} (\bibinfo {year} {2016})}\BibitemShut {NoStop}%
\bibitem [{\citenamefont {Garc\'{\i}a-Mata}\ \emph {et~al.}(2017)\citenamefont
  {Garc\'{\i}a-Mata}, \citenamefont {Giraud}, \citenamefont {Georgeot},
  \citenamefont {Martin}, \citenamefont {Dubertrand},\ and\ \citenamefont
  {Lemari\'e}}]{garciamata2017scaling}%
  \BibitemOpen
  \bibfield  {author} {\bibinfo {author} {\bibfnamefont {I.}~\bibnamefont
  {Garc\'{\i}a-Mata}}, \bibinfo {author} {\bibfnamefont {O.}~\bibnamefont
  {Giraud}}, \bibinfo {author} {\bibfnamefont {B.}~\bibnamefont {Georgeot}},
  \bibinfo {author} {\bibfnamefont {J.}~\bibnamefont {Martin}}, \bibinfo
  {author} {\bibfnamefont {R.}~\bibnamefont {Dubertrand}},\ and\ \bibinfo
  {author} {\bibfnamefont {G.}~\bibnamefont {Lemari\'e}},\ }\bibfield  {title}
  {\bibinfo {title} {Scaling theory of the {A}nderson transition in random
  graphs: Ergodicity and universality},\ }\href
  {https://doi.org/10.1103/PhysRevLett.118.166801} {\bibfield  {journal}
  {\bibinfo  {journal} {Phys. Rev. Lett.}\ }\textbf {\bibinfo {volume} {118}},\
  \bibinfo {pages} {166801} (\bibinfo {year} {2017})}\BibitemShut {NoStop}%
\bibitem [{\citenamefont {Sonner}\ \emph {et~al.}(2017)\citenamefont {Sonner},
  \citenamefont {Tikhonov},\ and\ \citenamefont
  {Mirlin}}]{sonner2017multifractality}%
  \BibitemOpen
  \bibfield  {author} {\bibinfo {author} {\bibfnamefont {M.}~\bibnamefont
  {Sonner}}, \bibinfo {author} {\bibfnamefont {K.~S.}\ \bibnamefont
  {Tikhonov}},\ and\ \bibinfo {author} {\bibfnamefont {A.~D.}\ \bibnamefont
  {Mirlin}},\ }\bibfield  {title} {\bibinfo {title} {Multifractality of wave
  functions on a {C}ayley tree: From root to leaves},\ }\href
  {https://doi.org/10.1103/PhysRevB.96.214204} {\bibfield  {journal} {\bibinfo
  {journal} {Phys. Rev. B}\ }\textbf {\bibinfo {volume} {96}},\ \bibinfo
  {pages} {214204} (\bibinfo {year} {2017})}\BibitemShut {NoStop}%
\bibitem [{\citenamefont {Biroli}\ and\ \citenamefont
  {Tarzia}(2018)}]{biroli2018delocalization}%
  \BibitemOpen
  \bibfield  {author} {\bibinfo {author} {\bibfnamefont {G.}~\bibnamefont
  {Biroli}}\ and\ \bibinfo {author} {\bibfnamefont {M.}~\bibnamefont
  {Tarzia}},\ }\href@noop {} {\bibinfo {title} {Delocalization and ergodicity
  of the {A}nderson model on {B}ethe lattices}} (\bibinfo {year} {2018}),\
  \Eprint {https://arxiv.org/abs/1810.07545} {arXiv:1810.07545
  [cond-mat.dis-nn]} \BibitemShut {NoStop}%
\bibitem [{\citenamefont {Kravtsov}\ \emph {et~al.}(2018)\citenamefont
  {Kravtsov}, \citenamefont {Altshuler},\ and\ \citenamefont
  {Ioffe}}]{kravtsov2018nonergodic}%
  \BibitemOpen
  \bibfield  {author} {\bibinfo {author} {\bibfnamefont {V.~E.}\ \bibnamefont
  {Kravtsov}}, \bibinfo {author} {\bibfnamefont {B.~L.}\ \bibnamefont
  {Altshuler}},\ and\ \bibinfo {author} {\bibfnamefont {L.~B.}\ \bibnamefont
  {Ioffe}},\ }\bibfield  {title} {\bibinfo {title} {Non-ergodic delocalized
  phase in anderson model on bethe lattice and regular graph},\ }\href
  {https://doi.org/https://doi.org/10.1016/j.aop.2017.12.009} {\bibfield
  {journal} {\bibinfo  {journal} {Annals of Physics}\ }\textbf {\bibinfo
  {volume} {389}},\ \bibinfo {pages} {148 } (\bibinfo {year}
  {2018})}\BibitemShut {NoStop}%
\bibitem [{\citenamefont {Tikhonov}\ and\ \citenamefont
  {Mirlin}(2019)}]{tikhonov2019critical}%
  \BibitemOpen
  \bibfield  {author} {\bibinfo {author} {\bibfnamefont {K.~S.}\ \bibnamefont
  {Tikhonov}}\ and\ \bibinfo {author} {\bibfnamefont {A.~D.}\ \bibnamefont
  {Mirlin}},\ }\bibfield  {title} {\bibinfo {title} {Critical behavior at the
  localization transition on random regular graphs},\ }\href
  {https://doi.org/10.1103/PhysRevB.99.214202} {\bibfield  {journal} {\bibinfo
  {journal} {Phys. Rev. B}\ }\textbf {\bibinfo {volume} {99}},\ \bibinfo
  {pages} {214202} (\bibinfo {year} {2019})}\BibitemShut {NoStop}%
\bibitem [{\citenamefont {Savitz}\ \emph {et~al.}(2019)\citenamefont {Savitz},
  \citenamefont {Peng},\ and\ \citenamefont {Refael}}]{savitz2019anderson}%
  \BibitemOpen
  \bibfield  {author} {\bibinfo {author} {\bibfnamefont {S.}~\bibnamefont
  {Savitz}}, \bibinfo {author} {\bibfnamefont {C.}~\bibnamefont {Peng}},\ and\
  \bibinfo {author} {\bibfnamefont {G.}~\bibnamefont {Refael}},\ }\bibfield
  {title} {\bibinfo {title} {Anderson localization on the {B}ethe lattice using
  cages and the {W}egner flow},\ }\href
  {https://doi.org/10.1103/PhysRevB.100.094201} {\bibfield  {journal} {\bibinfo
   {journal} {Phys. Rev. B}\ }\textbf {\bibinfo {volume} {100}},\ \bibinfo
  {pages} {094201} (\bibinfo {year} {2019})}\BibitemShut {NoStop}%
\bibitem [{\citenamefont {Garc\'{\i}a-Mata}\ \emph {et~al.}(2020)\citenamefont
  {Garc\'{\i}a-Mata}, \citenamefont {Martin}, \citenamefont {Dubertrand},
  \citenamefont {Giraud}, \citenamefont {Georgeot},\ and\ \citenamefont
  {Lemari\'e}}]{garciamata2020two}%
  \BibitemOpen
  \bibfield  {author} {\bibinfo {author} {\bibfnamefont {I.}~\bibnamefont
  {Garc\'{\i}a-Mata}}, \bibinfo {author} {\bibfnamefont {J.}~\bibnamefont
  {Martin}}, \bibinfo {author} {\bibfnamefont {R.}~\bibnamefont {Dubertrand}},
  \bibinfo {author} {\bibfnamefont {O.}~\bibnamefont {Giraud}}, \bibinfo
  {author} {\bibfnamefont {B.}~\bibnamefont {Georgeot}},\ and\ \bibinfo
  {author} {\bibfnamefont {G.}~\bibnamefont {Lemari\'e}},\ }\bibfield  {title}
  {\bibinfo {title} {Two critical localization lengths in the {A}nderson
  transition on random graphs},\ }\href
  {https://doi.org/10.1103/PhysRevResearch.2.012020} {\bibfield  {journal}
  {\bibinfo  {journal} {Phys. Rev. Research}\ }\textbf {\bibinfo {volume}
  {2}},\ \bibinfo {pages} {012020} (\bibinfo {year} {2020})}\BibitemShut
  {NoStop}%
\bibitem [{\citenamefont {Tarzia}(2020)}]{tarzia2020manybody}%
  \BibitemOpen
  \bibfield  {author} {\bibinfo {author} {\bibfnamefont {M.}~\bibnamefont
  {Tarzia}},\ }\bibfield  {title} {\bibinfo {title} {Many-body localization
  transition in hilbert space},\ }\href
  {https://doi.org/10.1103/PhysRevB.102.014208} {\bibfield  {journal} {\bibinfo
   {journal} {Phys. Rev. B}\ }\textbf {\bibinfo {volume} {102}},\ \bibinfo
  {pages} {014208} (\bibinfo {year} {2020})}\BibitemShut {NoStop}%
\bibitem [{\citenamefont {Biroli}\ and\ \citenamefont
  {Tarzia}(2017)}]{biroli2017delocalized}%
  \BibitemOpen
  \bibfield  {author} {\bibinfo {author} {\bibfnamefont {G.}~\bibnamefont
  {Biroli}}\ and\ \bibinfo {author} {\bibfnamefont {M.}~\bibnamefont
  {Tarzia}},\ }\bibfield  {title} {\bibinfo {title} {Delocalized glassy
  dynamics and many-body localization},\ }\href
  {https://doi.org/10.1103/PhysRevB.96.201114} {\bibfield  {journal} {\bibinfo
  {journal} {Phys. Rev. B}\ }\textbf {\bibinfo {volume} {96}},\ \bibinfo
  {pages} {201114} (\bibinfo {year} {2017})}\BibitemShut {NoStop}%
\bibitem [{\citenamefont {Biroli}\ and\ \citenamefont
  {Tarzia}(2020)}]{biroli2020anomalous}%
  \BibitemOpen
  \bibfield  {author} {\bibinfo {author} {\bibfnamefont {G.}~\bibnamefont
  {Biroli}}\ and\ \bibinfo {author} {\bibfnamefont {M.}~\bibnamefont
  {Tarzia}},\ }\bibfield  {title} {\bibinfo {title} {Anomalous dynamics on the
  ergodic side of the many-body localization transition and the glassy phase of
  directed polymers in random media},\ }\href
  {https://doi.org/10.1103/PhysRevB.102.064211} {\bibfield  {journal} {\bibinfo
   {journal} {Phys. Rev. B}\ }\textbf {\bibinfo {volume} {102}},\ \bibinfo
  {pages} {064211} (\bibinfo {year} {2020})}\BibitemShut {NoStop}%
\bibitem [{\citenamefont {Shirley}(1965)}]{shirley1965solution}%
  \BibitemOpen
  \bibfield  {author} {\bibinfo {author} {\bibfnamefont {J.~H.}\ \bibnamefont
  {Shirley}},\ }\bibfield  {title} {\bibinfo {title} {Solution of the
  {S}chr\"odinger equation with a hamiltonian periodic in time},\ }\href
  {https://doi.org/10.1103/PhysRev.138.B979} {\bibfield  {journal} {\bibinfo
  {journal} {Phys. Rev.}\ }\textbf {\bibinfo {volume} {138}},\ \bibinfo {pages}
  {B979} (\bibinfo {year} {1965})}\BibitemShut {NoStop}%
\bibitem [{\citenamefont {Sambe}(1973)}]{sambe1973steady}%
  \BibitemOpen
  \bibfield  {author} {\bibinfo {author} {\bibfnamefont {H.}~\bibnamefont
  {Sambe}},\ }\bibfield  {title} {\bibinfo {title} {Steady states and
  quasienergies of a quantum-mechanical system in an oscillating field},\
  }\href {https://doi.org/10.1103/PhysRevA.7.2203} {\bibfield  {journal}
  {\bibinfo  {journal} {Phys. Rev. A}\ }\textbf {\bibinfo {volume} {7}},\
  \bibinfo {pages} {2203} (\bibinfo {year} {1973})}\BibitemShut {NoStop}%
\bibitem [{\citenamefont {Pietracaprina}\ \emph {et~al.}(2016)\citenamefont
  {Pietracaprina}, \citenamefont {Ros},\ and\ \citenamefont
  {Scardicchio}}]{pietracaprina2016forward}%
  \BibitemOpen
  \bibfield  {author} {\bibinfo {author} {\bibfnamefont {F.}~\bibnamefont
  {Pietracaprina}}, \bibinfo {author} {\bibfnamefont {V.}~\bibnamefont {Ros}},\
  and\ \bibinfo {author} {\bibfnamefont {A.}~\bibnamefont {Scardicchio}},\
  }\bibfield  {title} {\bibinfo {title} {Forward approximation as a mean-field
  approximation for the {A}nderson and many-body localization transitions},\
  }\href {https://doi.org/10.1103/PhysRevB.93.054201} {\bibfield  {journal}
  {\bibinfo  {journal} {Phys. Rev. B}\ }\textbf {\bibinfo {volume} {93}},\
  \bibinfo {pages} {054201} (\bibinfo {year} {2016})}\BibitemShut {NoStop}%
\bibitem [{Note1()}]{Note1}%
  \BibitemOpen
  \bibinfo {note} {The Shirley Hamiltonian \protect \textup {\hbox
  {\mathsurround \z@ \protect \normalfont (\ignorespaces \ref
  {eq:H-shirley}\unskip \@@italiccorr )}} is translation-invariant in the
  Shirley direction and the initial state`s Shirley index can be chosen to be
  $n=0$ without loss of generality}\BibitemShut {NoStop}%
\bibitem [{Note2()}]{Note2}%
  \BibitemOpen
  \bibinfo {note} {Formally, $n_i^{(q)}=\DOTSB \sum@ \slimits@ _{i=1}^q
  s_i^{(q)}$ where $s_i^{(q)}=\pm 1,0$ is the change in the Shirley index on
  path $q$ at step $i$}\BibitemShut {NoStop}%
\bibitem [{Note3()}]{Note3}%
  \BibitemOpen
  \bibinfo {note} {The role of interferences of this kind were studied in a
  different context of transport in a driven one-dimensional Anderson
  insulator~\cite {localisation2017agarwal}.}\BibitemShut {Stop}%
\bibitem [{Note4()}]{Note4}%
  \BibitemOpen
  \bibinfo {note} {$K$ can be fractional as well. For instance, if a fraction
  $p$ and $1-p$ of the sites have $K_1$ and $K_2$ descendants respectively, the
  average branching number is $K=pK_1 + (1-p)K_2$.}\BibitemShut {Stop}%
\bibitem [{sup()}]{supp}%
  \BibitemOpen
  \href@noop {} {}\bibinfo {note} {See supplementary material at
  [URL].}\BibitemShut {Stop}%
\bibitem [{\citenamefont {Roy}\ and\ \citenamefont
  {Lazarides}(2020)}]{roy2020strong}%
  \BibitemOpen
  \bibfield  {author} {\bibinfo {author} {\bibfnamefont {S.}~\bibnamefont
  {Roy}}\ and\ \bibinfo {author} {\bibfnamefont {A.}~\bibnamefont
  {Lazarides}},\ }\bibfield  {title} {\bibinfo {title} {Strong ergodicity
  breaking due to local constraints in a quantum system},\ }\href
  {https://doi.org/10.1103/PhysRevResearch.2.023159} {\bibfield  {journal}
  {\bibinfo  {journal} {Phys. Rev. Research}\ }\textbf {\bibinfo {volume}
  {2}},\ \bibinfo {pages} {023159} (\bibinfo {year} {2020})}\BibitemShut
  {NoStop}%
\bibitem [{\citenamefont {Agarwal}\ \emph {et~al.}(2017)\citenamefont
  {Agarwal}, \citenamefont {Ganeshan},\ and\ \citenamefont
  {Bhatt}}]{localisation2017agarwal}%
  \BibitemOpen
  \bibfield  {author} {\bibinfo {author} {\bibfnamefont {K.}~\bibnamefont
  {Agarwal}}, \bibinfo {author} {\bibfnamefont {S.}~\bibnamefont {Ganeshan}},\
  and\ \bibinfo {author} {\bibfnamefont {R.~N.}\ \bibnamefont {Bhatt}},\
  }\bibfield  {title} {\bibinfo {title} {Localization and transport in a
  strongly driven {A}nderson insulator},\ }\href
  {https://doi.org/10.1103/PhysRevB.96.014201} {\bibfield  {journal} {\bibinfo
  {journal} {Phys. Rev. B}\ }\textbf {\bibinfo {volume} {96}},\ \bibinfo
  {pages} {014201} (\bibinfo {year} {2017})}\BibitemShut {NoStop}%
\end{thebibliography}%


\clearpage
{
\onecolumngrid
\begin{center}
\textbf{Supplementary material: How periodic driving stabilises and destabilises Anderson localisation on random trees}\\

\end{center}
\bigskip
}

\twocolumngrid

\setcounter{equation}{0}
\renewcommand{\theequation}{S\arabic{equation}}
\setcounter{figure}{0}
\renewcommand{\thefigure}{S\arabic{figure}}
\setcounter{page}{1}
\renewcommand{\thepage}{S\arabic{page}}

\subsection{Transfer matrix for FSA in Shirley picture}
We briefly describe the transfer matrix implementation of the FSA in the Shirley picture for a single physical path $\alpha_0\shortrightarrow\alpha_1\shortrightarrow\cdots\alpha_d$.
The transfer matrix acts on a space of $(d+1)\times (2d+1)$ sites on the Shirley graph as there are $d+1$ physical sites and $2d+1$ Shirley rungs $n\in[-d,d]$.
We first introduce a matrix $\mathcal{A}$ proportional to the adjacency matrix of the \emph{directed} Shirley graph.
The directed-ness of the graph stems from the fact that amplitude from site $(\alpha_i,n)$ can only be transported to $(\alpha_{i+1},n+s)$ with $s=\pm1,0$ and not back. Formally, 
\eq{
\mathcal{A} =\Gamma \sum_{i=0}^{d-1}\sum_{n=-d}^d\sum_{s=\pm1,0}\ket{\alpha_{i+1},n+s}\bra{\alpha_i,n}\,.
}
Additionally, we introduce a diagonal matrix defined on the vertices of the graph as 
\eq{
\mathcal{D} = \sum_{i=0}^d\sum_{n=-d}^d(\epsilon_{\alpha_i}+n\omega-\epsilon_{\alpha_0})^{-1}\ket{\alpha_i,n}\bra{\alpha_i,n}\,.
}
The transfer matrix then given by 
\eq{
\mathcal{T}= \mathcal{D}\mathcal{A}\,,
}
such that for the initial state $\ket{\psi_0}=\ket{\alpha_0,0}$, we have
\eq{
\mathcal{T}^x\ket{\psi_0}=\sum_{n=-x}^x\psi_{\alpha_x,n}\ket{\alpha_x,n}\,
}
where $\psi_{\alpha_x,n}$'s are the FSA amplitudes of interest.

\subsection{FSA amplitudes at high-frequency}
In this section, we present the details of the derivation of Eq.~\ref{eq:Lambda-d-highfreq} at leading order.
Since, the Shirley Hamiltonian \eqref{eq:H-shirley} is invariant under $\omega\to-\omega$, the leading order correction is expected to be $\mathcal{O}(1/\omega^2)$; hence we consider all the paths that deviate from the static path at at most two sites.
As a matter of notation, the amplitude  of a path that deviates from the static path at site $\alpha_i$, such that $n_{\alpha_i}^{(q)}=\pm 1$ will be denoted as $w_{i_\pm}$. 
The paths that deviate at exactly two sites from the static path are also forced to have $n_{\alpha_i}^{(q)}=\pm 1$ except for the two paths where $n_{\alpha_{d-1}}^{(q)}=\pm 1$ and $n_{\alpha_d}^{(q)}=\pm 2$. 
We denote the amplitude of paths that deviate at sites $\alpha_i$ and $\alpha_j$ with $n_{\alpha_i}^{(q)}=\pm 1$ and $n_{\alpha_j}^{(q)}=\pm 1$ as $w_{i_\pm j_\pm}$. Additionally, we denote the amplitudes of the two paths with $n_{\alpha_{d-1}}^{(q)}=\pm 1$ and $n_{\alpha_d}^{(p)}=\pm 2$ as $w_+^{(2)}$ and ${w}_-^{(2)}$ respectively. With this notation, to $\mathcal{O}(1/\omega^2)$, $G=(\sum_{q}w_q)^2$ can be written as 
\begin{subequations}
\begin{align}
G =& w_\mathrm{s}^2 +  (\sum_{i=1}^d w_{i_+})^2 + (\sum_{i=1}^d w_{i_-})^2  \label{eq:Gd1}\\
&+2 w_\mathrm{s}\sum_{i=1}^d(w_{i_+}+w_{i_-}) +2\sum_{i,j}w_{i_+}w_{j_-}\label{eq:Gd2}\\
&+2w_\mathrm{s}\left(\sum_{i,j}\sum_{\eta,\eta^\prime=\pm}w_{i_\eta j_{\eta^\prime}}+w_+^{(2)}+w_-^{(2)}\right).
\label{eq:Gd3}
\end{align}
\label{eq:G-det}
\end{subequations}
Using $\Delta_i\equiv\epsilon_i-\epsilon_0$, the amplitudes $w_{i_\pm}$ and $w_{i_\pm j_\pm}$ can be related to $w_\mathrm{s}$ as 
\begin{equation}
w_{i_\pm} = w_\mathrm{s}\frac{\Delta_i}{\Delta_i\pm\omega},~w_{i_\pm j_\pm} = w_\mathrm{s}\frac{\Delta_i\Delta_j}{(\Delta_i\pm\omega)(\Delta_j\pm\omega)},
\label{eq:w12-amp}
\end{equation}
and 
\begin{equation}
w^{(2)}_\pm = w_\mathrm{s}\frac{\Delta_{d-1}\Delta_d}{(\Delta_{d-1}\pm\omega)(\Delta_{d}\pm2\omega)}.
\label{eq:w2-amp}
\end{equation}
Using Eqs.~\ref{eq:w12-amp} and \ref{eq:w2-amp}, and $\omega\gg\Delta_i$ in Eq.~\ref{eq:G-det}, we have
\begin{subequations}
\begin{align}
G =& w_\mathrm{s}^2 +  2w_\mathrm{s}^2\sum_{i=1}^d \frac{\Delta_i^2}{\omega^2} + 2w_\mathrm{s}^2\sum_{i\neq j}\frac{\Delta_i\Delta_j}{\omega^2} \label{eq:Gl1} \\
&-4w_\mathrm{s}^2\sum_{i=1}^d\frac{\Delta_i^2}{\omega^2} -2w_\mathrm{s}^2\sum_{i,j}\frac{\Delta_i\Delta_j}{\omega^2}\label{eq:Gl2}\\
&+w_\mathrm{s}^2\frac{\Delta_{d-1}\Delta_{d}}{\omega^2}\label{eq:Gl3}.
\end{align}
\label{eq:G-exp}
\end{subequations}
The terms in Eqs.~\ref{eq:Gl1}, \ref{eq:Gl2}, and \ref{eq:Gl3} correspond exactly to their counterparts in Eqs.~\ref{eq:Gd1},\ref{eq:Gd2}, and \ref{eq:Gd3} respectively (The summation in Eq.~\ref{eq:Gd3} vanishes.). Equation~\ref{eq:G-exp} can be reorganised to yield
\begin{equation}
G = w_\mathrm{s}^2\left(1-\frac{4}{\omega^2}\sum_{i=1}^d\Delta_i^2+\frac{\Delta_{d-1}\Delta_{d}}{\omega^2}\right),
\end{equation}
which directly leads to Eq.~\ref{eq:Lambda-d-highfreq}.

\subsection{More on the statistics of Shirley path amplitudes}
\begin{figure}
    \includegraphics[width=\linewidth]{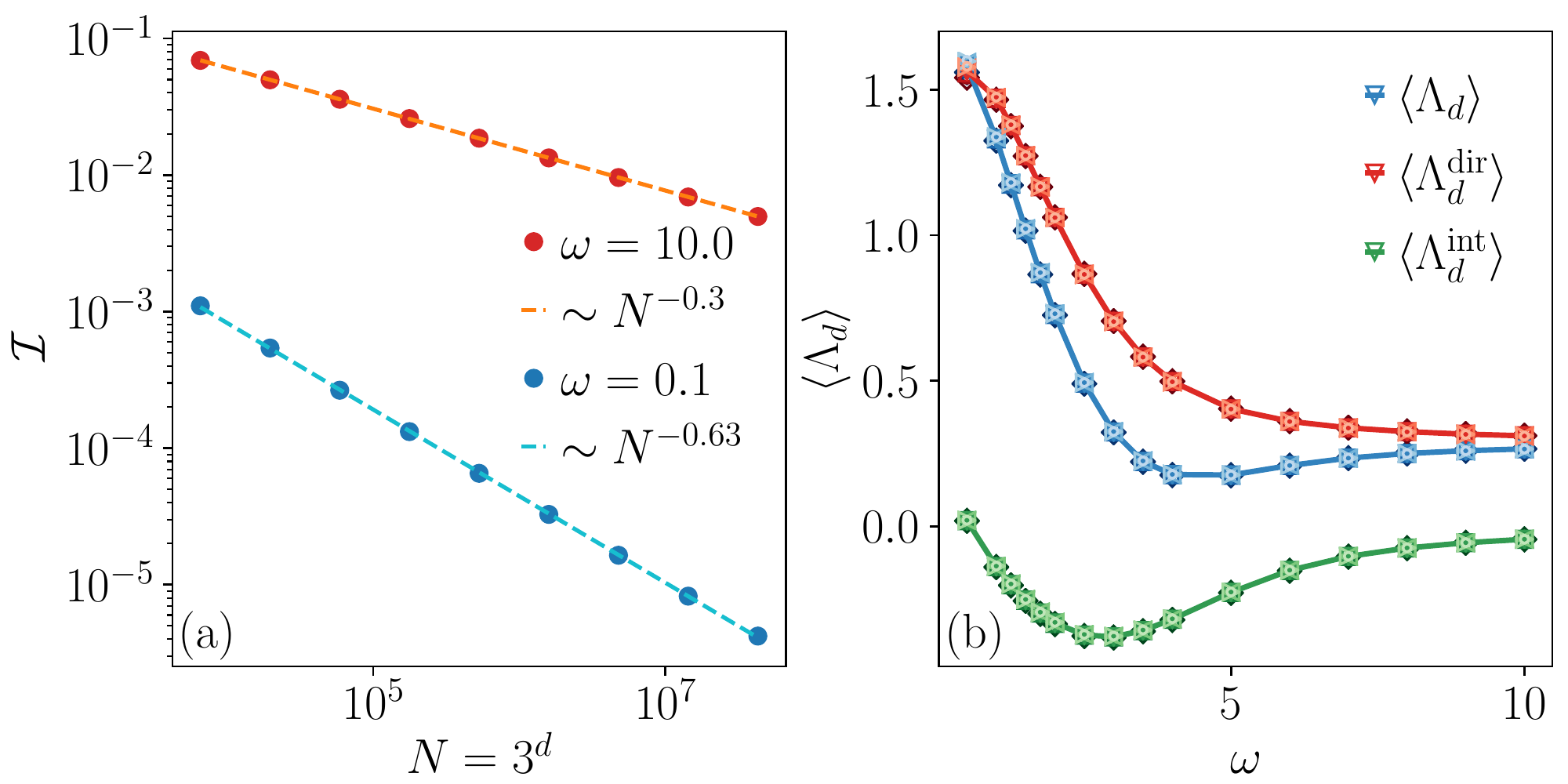}
    \caption{(a) The IPR defined in Eq.~\ref{eq:IPR}. The scaling with $N=3^d$ shows that with decreasing $\omega$, qualitatively more paths contribute to the total amplitude. (b) The behaviour of $\braket{\Lambda_d}$ (blue), $\braket{\Lambda_d^\mathrm{dir}}$ (red), and $\braket{\Lambda_d^\mathrm{int}}$ (green) as a function of $\omega$. The different markers and colour-intensities show $d=8$ to $d=16$, all of which are well converged with $d$.}
    \label{fig:lambda-dir-cross-ipr}
\end{figure}
In this section, we present some more results on the statistics of Shirley path amplitudes. 
We first show evidence that with decreasing $\omega$, the number of paths that contribute to the FSA amplitudes increases qualitatively.
To this end, we define an effective inverse participation ratio
\eq{
\mathcal{I} = \sum_{q}\left(\frac{\vert w_q\vert}{\sum_{q^\prime} \vert w_{q^\prime}\vert}\right)^2\,.
\label{eq:IPR}
}
In the limit of $\omega\to 0$, when there is only one path (the static path) with a finite amplitude, we expect $\mathcal{I}\sim N^0$ with $N=3^d$ the total number of paths.
On the other hand, in the limit of $\omega\to 0$, where all the Shirley paths are equivalent, we expect $\mathcal{I}\sim N^{-1}$.
The results shown in Fig.~\ref{fig:lambda-dir-cross-ipr}(a) suggest that $\mathcal{I}\sim N^{-\alpha} $ with $\alpha$ growing with decreasing $\omega$. 
This implies that the total FSA amplitude is more and more delocalised over all the Shirley paths as $\omega$ is decreased.

In the main text, we had decomposed $\Lambda_d = \Lambda_d^\mathrm{dir}+\Lambda_d^\mathrm{int}$, where the first term is the direct term and the second term encodes the interferences between the paths. 
In Fig.~\ref{fig:lambda-dir-cross-ipr}(b), we show the behaviour of $\braket{\Lambda_d}$, $\braket{\Lambda_d^\mathrm{dir}}$, and $\braket{\Lambda_d^\mathrm{int}}$ with $\omega$.
With increasing $\omega$, since fewer paths contribute and the strength of each resonance also weakens, $\braket{\Lambda_d^\mathrm{dir}}$ decreases monotonically.
On the other hand, $\braket{\Lambda_d^\mathrm{int}}$ is a non-monotonic function of $\omega$, which leads to the eventual non-monotonicity in $\Gamma_c$ as a function of $\omega$ as explain in the main text.
Note that for $\omega/W\ll 1$, $\vert\braket{\Lambda_d^\mathrm{int}}\vert \ll \vert\braket{\Lambda_d^\mathrm{dir}}\vert$ and also $\braket{\Lambda_d^\mathrm{int}} <0$ which justifies approximating $\Lambda_d\approx\Lambda_d^\mathrm{dir}$ in the low-frequency regime.

\subsection{Effective number of paths in low-frequency regime}
We show, in this section, that $\lim_{d\to\infty}\ln(N_q^\mathrm{eff})/2d =c$ where $c$ is a constant and $N_q^\mathrm{eff}$ is the total length of all paths spent inside the resonant strip, $\vert n \vert <W/\omega$.
We compute $N_q^\mathrm{eff}$ as follows. At distance $x$, let us the denote the probability that a path passes through rung $n_x$ as $\mathcal{P}_{n_x}$, which is given by the trinomial distribution
\eq{
\begin{split}
\mathcal{P}_{n_x}(n) &= \sum_{a,b=0}^x\frac{\delta_{a-b,n}}{3^x}\frac{x!}{a!~b!~(x-a-b)!}\\
&\overset{x\gg 1}{\approx} \frac{1}{\sqrt{2\pi\sigma^2}}e^{-n^2/2\sigma^2};~~~\sigma^2 = 2x/3\,.
\end{split}
}
The total probability that a path is inside the strip of resonances at distance $x$ is given by 
\eq{
m_x=\int_{-W/\omega}^{W/\omega}dn~\mathcal{P}_{n_x}(n) \approx \mathrm{Erf}\left[\frac{W}{\omega}\sqrt{\frac{3}{2x}}\right]\,.
}
Since, the total number of paths is $3^d$, the effective number of paths $N_q^\mathrm{eff}$ is
\eq{
N_q^\mathrm{eff} = 3^d \frac{1}{d}\int_0^d dx~m_x\,,
\label{eq:neff}
}
as such in the limit of $d\to\infty$, we have 
\begin{equation}
\lim_{d\to\infty}\frac{1}{2d}\ln N_q^\mathrm{eff} = \ln\sqrt{3}\equiv c\,.
\label{eq:neff1}
\end{equation}

\end{document}